\documentclass[superscriptaddress,prd,aps,showpacs,nofootinbib,showkeywords,eqsecnum]{revtex4-2}

\usepackage{graphicx}
\usepackage{amsmath,amssymb}
\allowdisplaybreaks
\usepackage{hyperref}
\usepackage{xcolor}
\usepackage{caption}
\usepackage{float}
\usepackage{verbatim}
\usepackage{subcaption}
\usepackage{array,multirow}

\begin{document}

	\title{Power-law Bianchi type I inflation with multiple vector fields}
	
	\author{Duy H. Nguyen}
	\email{duy.nguyenhoang@phenikaa-uni.edu.vn}
	\affiliation{Phenikaa Institute for Advanced Study, Phenikaa University, Hanoi 12116, Vietnam}	
	
	\author{Tuan Q. Do }
	\email{tuan.doquoc@phenikaa-uni.edu.vn}
	\affiliation{Phenikaa Institute for Advanced Study, Phenikaa University, Hanoi 12116, Vietnam}
	
	\begin{abstract}
		We investigate an inflationary anisotropic universe in a supergravity-motivated model with one scalar field non-minimally coupled to multiple vector fields. We restrict ourselves to the Bianchi type I metric, which describes a homogeneous but anisotropic universe. For consistency, we consider a configuration consisting of one homogeneous scalar field and three mutually orthogonal vector fields. As a result, we find four types of power-law solutions, classified according to the number of non-vanishing vector fields. Moreover, we show that all these solutions are stable under certain conditions on the model parameters, thereby defining stability regions described both quantitatively and qualitatively. Interestingly, our analysis suggests that vector fields with significantly larger coupling constants tend to persist as the universe expands, while those with significantly smaller coupling constants are eventually diluted. On the other hand, we also find that the anisotropies depend on the number of persisting vector fields. Furthermore, our claims are confirmed by numerical calculations. This work may therefore shed light on how vector fields and anisotropies evolve in an inflationary universe.
	\end{abstract}
	
	%\pacs{Insert PACS numbers here}
	
	\maketitle
		
\section{Introduction} \label{sec1}
In modern cosmology, the cosmological principle, which is a hypothesis stating that the universe is homogeneous and isotropic on large scales, has served as an underlying assumption. A universe obeying the cosmological principle is uniquely described by the Friedmann-Lemaitre-Robertson-Walker (FLRW) metric \cite{Aluri:2022hzs}. Moreover, the cosmological principle is supported by the cosmic no-hair conjecture proposed by Hawking et al. \cite{Gibbons:1977mu,Hawking:1981fz}. According to the conjecture, all inhomogeneities and anisotropies, a.k.a. classical spatial hairs, vanish as the universe expands. Therefore, if the conjecture holds, the initial inhomogeneities and anisotropies become unimportant, and the cosmological principle is realized at later times. Remarkably, the validity of the cosmic no-hair conjecture has been examined extensively by many authors, see Refs. \cite{Wald:1983ky,Barrow:1987ia,Mijic:1987bq,Kitada:1991ih,Kleban:2016sqm,East:2015ggf,Carroll:2017kjo,Kaloper:1991rw,Barrow:2005qv,Tahara:2018orv,Starobinsky:2019xdp,Galeev:2021xit,Nojiri:2022idp,Starobinsky:1982mr,Muller:1989rp,Barrow:1984zz,Jensen:1986nf,SteinSchabes:1986sy} for an incomplete list. Many of these works have focused on the inflationary phase of the early universe, which was proposed as a resolution to several fundamental problems in modern cosmology such as the flatness, horizon and magnetic monopole problems \cite{Starobinsky:1980te,Guth:1980zm,Linde:1981mu}, to assess whether the cosmic no-hair conjecture is valid.

Recently, the cosmological principle and the cosmic no-hair conjecture have been challenged by observations of the Cosmic Microwave Background (CMB). In particular, the Wilkinson Microwave Anisotropy Probe (WMAP) detected two anomalies in the CMB, including the cold spot and hemispheric asymmetry \cite{WMAP:2012nax}, which were also confirmed by the Planck satellite \cite{Planck:2019evm}. Therefore, it is reasonable to seek an exotic inflationary model that violates the cosmological principle and the cosmic no-hair conjecture \cite{Schwarz:2015cma}, since these CMB anomalies are believed not to arise in standard inflationary models, in which the cosmological principle holds and the cosmic no-hair conjecture is not violated. One of the candidates for the origin of the observed anomalies is vector fields. However, vector fields are usually ignored in studies of inflation since they are believed to decay very quickly as the universe expands, which is consistent with the cosmic no-hair conjecture. 

However, an anisotropic inflationary model proposed by Kanno, Soda, and Watanabe (KSW) showed the opposite behavior \cite{Watanabe:2009ct,Kanno:2010nr}. Motivated by supergravity, they introduced a non-minimal coupling between one scalar field $\phi$, of which the potential is $V(\phi)\propto \exp(\lambda\phi)$, and one vector field $A_\mu$, with a gauge kinetic term of the form $f^2(\phi)F^{\mu\nu}F_{\mu\nu} \propto \exp(2\rho \phi) F^{\mu\nu}F_{\mu\nu}$. It is worth noting that in the KSW model, the spacetime is chosen to be a locally rotationally symmetric (LRS) Bianchi type I metric, which has the form,
\begin{align}
ds^2&=-dt^2+a^2(t)dx^2+b^2(t)(dy^2+dz^2).
\end{align}
Moreover, in order to be compatible with the metric, the field configuration consisting of one scalar field $\phi$ and one vector field $A_\mu$,
\begin{align}\label{KSW field configuration}
\phi=\phi(t),\quad A_\mu=(0,A_a(t),0,0),
\end{align}
is chosen. Here, it is assumed that our universe has a privileged direction aligned along the $x$-axis. As a result, they succeeded in deriving an anisotropic power-law inflationary solution, which has been shown to be stable against perturbations under some specific conditions on the parameters. In fact, the gauge kinetic function $f(\phi)$ plays a crucial role in keeping the anisotropies from being diluted. The KSW model therefore becomes a true counterexample to the cosmic no-hair conjecture. Furthermore, CMB imprints of such an anisotropic inflation model have been studied extensively in Refs. \cite{Watanabe:2010fh,Dulaney:2010,Gumrukcuoglu:2010yc,Watanabe:2010bu,Bartolo:2012sd,Ohashi:2013qba,Soda:2012zm,Maleknejad:2012fw}. Interestingly, many other counterexamples, which are nothing but non-trivial extensions of the KSW model, have been proposed in recent years \cite{Emami:2010rm,Murata:2011wv,Bhowmick:2011em,Do:2011zza,Hervik:2011xm,Do:2011zz,Maleknejad:2012as,Thorsrud:2012mu,Yamamoto:2012tq,Ohashi:2013mka,Ohashi:2013pca,Ito:2015sxj,Do:2016ofi,Lahiri:2016jqv,Tirandari:2017nzy,Do:2017rva,Do:2017qyd,Ito:2017bnn,Holland:2017cza,Do:2017onf,Tirandari:2018mgf,Do:2020hjf,Do:2021lyf,Chen:2021nkf,Nguyen:2021emx,Do:2021pqk,Goodarzi:2022wli,Kanno:2022flo,Do:2023mqe,Pham:2023evo,Do:2025xhm}. 

An extension of the KSW model for multiple vector fields has been investigated in Ref. \cite{Yamamoto:2012tq}. By numerical methods, it was shown that the universe has the tendency to minimize the anisotropy as it expands. However, its exact analytical solutions have remained unsolved mostly due to the complexity of the system. In this paper, in order to derive the exact solutions we impose the Bianchi type I metric in its most general form,
\begin{align}\label{original BI metric}
ds^2=-dt^2+a^2(t) dx^2+b^2(t) dy^2+c^2(t) dz^2,
\end{align}
which includes the LRS one shown above as a special case. In accordance with the metric, it is reasonable to choose a simple field configuration that consists of one homogeneous scalar field and three orthogonal vector fields aligned along the three spatial axes as follows:
\begin{equation}\label{field configuration}
\begin{aligned}
\phi=\phi(t),\quad A_\mu=(0,A_a(t),0,0),\quad B_\mu=(0,0,B_b(t),0),\quad C_\mu=(0,0,0,C_c(t)).
\end{aligned}
\end{equation}
Of course, one can instead introduce a more complicated version of \eqref{field configuration} with multiple scalar fields and more than three vector fields, as long as it is consistent with the Bianchi type I metric. However, we choose the configuration \eqref{field configuration} since it represents the simplest realization. If the three vectors have equal magnitudes, i.e. $A_a(t)=B_b(t)=C_c(t)$, we obtain a configuration called the ``cosmic triad'', which has been studied previously. In Ref. \cite{Armendariz-Picon:2004say}, Armendáriz-Picón proposed a dark energy model in which a cosmic triad is responsible for the accelerated expansion of the present universe. On the other hand, an inflationary model driven by a cosmic triad was investigated by  Golovnev,  Mukhanov, and Vanchurin in Ref. \cite{Golovnev:2008cf}. It appears that the cosmic triad has received a lot of attention in recent years, see Refs. \cite{Wei:2006tn,Wei:2006ut,Chiba:2008eh,Zhang:2009tj,Golovnev:2009ks,Himmetoglu:2009qi,Zhang:2009yu,Yamamoto:2012sq,Funakoshi:2012ym,Setare:2013kja,Darabi:2014aaa,Landim:2016dxh,Oliveros:2019zkl,Gorji:2020vnh,Orjuela-Quintana:2020klr,Murata:2021vnb,Rodriguez-Benites:2023otm,Coelho:2025vmo} for an incomplete list of works on this configuration. It is important to note that the cosmic triad has been chosen in these works to guarantee the spatial isotropy of the universe. For the field configuration shown in Eq. \eqref{field configuration}, which will be considered in the present paper, the spatial isotropy of the universe has been generally assumed to be broken down.

After setting the background, we aim to derive all exact isotropic and anisotropic solutions of the proposed model and investigate their stability. Remarkably, our analysis reveals interesting behaviors of the vector fields and the anisotropies. It turns out that the fate of vector fields and anisotropies does not depend solely on the relative values of the coupling constants $\rho_i$ and $\lambda$, as in the KSW model, but also on the relative values of the coupling constants among themselves. In particular, as long as at least one of the coupling constants is sufficiently large, the vector field with the largest coupling constant $\rho_\text{max}$ and any vector field with a coupling constant that is smaller but sufficiently close to $\rho_\text{max}$ persist as the universe expands. However, any vector field with a coupling constant that is significantly smaller than $\rho_\text{max}$ is eventually diluted. Moreover, the fate of anisotropies also depends on the number of persisting vector fields. The results obtained in our paper may therefore shed light on how vector fields and anisotropies evolve in an inflationary universe.

Our paper will be organized as follows: (i) A brief introduction of our study has been written in Sect. \ref{sec1}. (ii) In Sect. \ref{sec2}, we will present the general setup of our model and derive the corresponding field equations. (iii) In Sect. \ref{sec3}, we will solve the field equations to find the anisotropic power-law solutions. (iv) In Sect. \ref{sec4}, we will investigate the stability of the solutions by transforming the field equations into autonomous equations of dynamical variables. Numerical calculations to support the above stability analysis will be conducted in this section. (v) Conclusions and further discussions will be given in Sect. \ref{sec5}. (vi) Finally, Appendices \ref{Appendix 1} and \ref{RH criterion} will be devoted to some detailed calculations.
%%%%%%%%%%%%%
\section{Model setup} \label{sec2}
We consider the Bianchi type I (BI) metric, which describes a homogeneous but anisotropic universe, as \eqref{original BI metric}, where $a(t)$, $b(t)$, and $c(t)$ are the scale factors corresponding to the $x$-, $y$-, and $z$-axes, respectively. It is convenient to parameterize the metric as
\begin{align}
ds^2=-dt^2+e^{2\alpha(t)-2\sigma_b(t)-2\sigma_c(t)}dx^2+e^{2\alpha(t)+2\sigma_b(t)}dy^2+e^{2\alpha(t)+2\sigma_c(t)}dz^2. \label{BI metric}
\end{align}
Note that there is another way to parameterize the metric \eqref{original BI metric}, which can be found in Ref. \cite{Soda:2012zm}.  It is also convenient to define the average Hubble parameter as
\begin{align}
H\equiv\frac{1}{3}\left(H_a+H_b+H_c\right)=\dot{\alpha},
\end{align}
where $H_a\equiv\dot{a}/a = \dot\alpha-\dot\sigma_b-\dot\sigma_c$, $H_b\equiv\dot{b}/b = \dot\alpha+\dot\sigma_b$, and $H_c\equiv\dot{c}/c =\dot\alpha +\dot\sigma_c$. In addition, we  define spatial anisotropies as follows
\begin{equation}\label{anysotropies along axes}
\begin{aligned}
X_b&\equiv\frac{H_b-H}{H}=\frac{\dot{\sigma}_b}{\dot{\alpha}}, \\
 X_c&\equiv\frac{H_c-H}{H}=\frac{\dot{\sigma}_c}{\dot{\alpha}}, \\  X_a&\equiv\frac{H_a-H}{H}=\frac{-\dot{\sigma}_b-\dot{\sigma}_c}{\dot{\alpha}}=-X_b-X_c,
\end{aligned}
\end{equation}
which measure the corresponding spatial deviations from isotropy along the $y$-, $z$-, and $x$-axes, respectively. Note that when two of the anisotropies are equal, say $X_b=X_c$, we can use the coordinate transformation $z\rightarrow c(t)z/b(t)$ to equalize the scale factors $b(t)$ and $c(t)$ and then recover the Bianchi type I metric that is locally rotationally symmetric about the $x$-axis (denoted by the $x$-rsBI metric for short) and has the following form,
\begin{align}\label{rotational symmetric BI metric}
ds^2&=-dt^2+a^2(t) dx^2+b^2(t)(dy^2+dz^2)=-dt^2+e^{2\alpha(t)-4\sigma(t)}dx^2+e^{2\alpha(t)+2\sigma(t)}(dy^2+dz^2),
\end{align}
where $\sigma(t)\equiv \sigma_b(t)=\sigma_c(t)$. Similarly, when $X_a=X_b=X_c$ (and consequently they are all equal to zero), we obtain the well-known FLRW metric as
\begin{align}\label{FLRW metric}
ds^2&=-dt^2+a^2(t)(dx^2+dy^2+dz^2)=-dt^2+e^{2\alpha(t)}(dx^2+dy^2+dz^2).
\end{align}

In many studies on anisotropic inflation, the metric has been taken to be that shown in Eq. \eqref{rotational symmetric BI metric}. A typical example is the KSW model \cite{Watanabe:2009ct,Kanno:2010nr}, in which the authors considered a supergravity-motivated action,
\begin{align}\label{one vector field action}
S=\int d^4x\sqrt{-g}\left[\frac{R}{2}-\frac{1}{2}\partial_\mu\phi\partial^\mu\phi-V(\phi)-\frac{1}{4} f^2(\phi)F_{\mu\nu}F^{\mu\nu}\right],
\end{align}
where the reduced Planck mass has been set to be one, i.e., $M_p=1$, for convenience. In addition,  $R$ is the Ricci scalar, $\phi$ is a scalar field, $V(\phi)$ is the potential, and $F_{\mu\nu}\equiv \partial_\mu A_\nu-\partial_\nu A_\mu$ is the field strength of the vector field $A_\mu$. In harmony with the metric \eqref{rotational symmetric BI metric}, a homogeneous scalar field and a vector field of the form \eqref{KSW field configuration} are introduced. The metric \eqref{rotational symmetric BI metric} has also appeared in other studies, e.g., Refs. \cite{Do:2011zz,Ohashi:2013pca,Holland:2017cza,Nguyen:2021emx,Do:2016ofi,Do:2020hjf}, with different actions and field configurations that are also required to be in harmony with the metric.

In our model, we would like to consider a more general action given by
\begin{align}\label{action}
S=\int d^4x\sqrt{-g}\left[\frac{R}{2}-\frac{1}{2}\partial_\mu\phi\partial^\mu\phi-V(\phi)-\frac{1}{4}\sum_{i=a,b,c} f_i^2(\phi){F_{(i)}}_{\mu\nu}{F_{(i)}}^{\mu\nu}\right],
\end{align}
which is a straightforward generalization of the action \eqref{one vector field action} to the case of three vector fields shown in \eqref{field configuration}. Here, it is understood that ${F_{(a)}}_{\mu\nu} \equiv \partial_\mu A_\nu -\partial_\nu A_\mu$, ${F_{(b)}}_{\mu\nu} \equiv \partial_\mu B_\nu -\partial_\nu B_\mu$, and ${F_{(c)}}_{\mu\nu} \equiv \partial_\mu C_\nu -\partial_\nu C_\mu$. From the principle of least action, we derive the corresponding Einstein field equations to be
\begin{align}\label{Einstein eq}
R_{\mu\nu}-\frac{1}{2}R g_{\mu\nu}=\partial_\mu\phi\partial_\nu\phi-g_{\mu\nu}\left[\frac{1}{2}\partial_\rho\phi\partial^\rho\phi+V(\phi)+\frac{1}{4}\sum_{i=a,b,c} f_i^2(\phi){F_{(i)}}_{\rho\sigma}{F_{(i)}}^{\rho\sigma}\right]+\sum_{i=a,b,c} f_i^2(\phi){F_{(i)}}_{\mu\sigma}{F_{(i)}}_{\nu}^{\;\;\sigma},
\end{align}
along with the equation of motion for the scalar field $\phi$ given by
\begin{align}\label{general scalar field eq}
\Box\phi-V'(\phi)-\frac{1}{2}\sum_{i=a,b,c} f_i(\phi)f_i'(\phi){F_{(i)}}_{\rho\sigma}{F_{(i)}}^{\rho\sigma}=0,
\end{align}
where the prime stands for a derivative with respect to $\phi$, and $\Box\equiv\frac{1}{\sqrt{-g}}\partial_\mu(\sqrt{-g}\partial^\mu)$ is the d'Alembert operator. Lastly, the corresponding equations of motion for three vector fields are found to be 
\begin{align}\label{general vector fields eq}
\partial_\mu\left[\sqrt{-g}f_i^2(\phi){F_{(i)}}^{\mu\nu}\right]=0,\quad\text{for }i=a,b,c.
\end{align}

Instead of the metric \eqref{rotational symmetric BI metric}, we consider the Bianchi type I metric in its most general form as shown in Eq. \eqref{BI metric}. In harmony with the chosen metric, we choose a configuration with one scalar field and three orthogonal vector fields aligned along three axes as displayed in Eq. \eqref{field configuration}. If we instead choose the KSW's field configuration \eqref{KSW field configuration}, the difference between the anisotropies in the $y$- and $z$-axes will decay quickly as the universe expands and therefore it is more convenient to consider the metric \eqref{rotational symmetric BI metric} instead of \eqref{BI metric} \cite{Soda:2012zm}. However, as we will see later, this is not the case for the field configuration \eqref{field configuration}. 

For the assumed metric \eqref{BI metric} and field configuration \eqref{field configuration}, non-trivial solutions for the three vector fields can be obtained from their equations of motion \eqref{general vector fields eq} as
\begin{align}\label{vector fields eq}
\dot{A}_a&=p_a f_a^{-2}e^{-\alpha-2\sigma_b-2\sigma_c},\quad\dot{B}_b=p_b f_b^{-2}e^{-\alpha+2\sigma_b},\quad\dot{C}_c=p_c f_c^{-2}e^{-\alpha+2\sigma_c},
\end{align}
where the dot stands for a derivative with respect to cosmic time $t$. In addition, $p_a$, $p_b$, and $p_c$ are constants of integration. Thanks to these solutions, we can therefore write explicitly the Einstein field equations \eqref{Einstein eq} as follows
\begin{align}
3\dot{\alpha}^2-\dot{\sigma}_b^2-\dot{\sigma}_c^2-\dot{\sigma}_b\dot{\sigma}_c&=\frac{1}{2}\dot{\phi}^2+V(\phi)+\frac{p_a^2}{2}f_a^{-2}(\phi)e^{-4\alpha-2\sigma_b-2\sigma_c}+\frac{p_b^2}{2}f_b^{-2}(\phi)e^{-4\alpha+2\sigma_b}+\frac{p_c^2}{2}f_c^{-2}(\phi)e^{-4\alpha+2\sigma_c},\label{Friedmann eq 1}\\
\ddot{\alpha}+3\dot{\alpha}^2&=V(\phi)+\frac{p_a^2}{6}f_a^{-2}(\phi)e^{-4\alpha-2\sigma_b-2\sigma_c}+\frac{p_b^2}{6}f_b^{-2}(\phi)e^{-4\alpha+2\sigma_b}+\frac{p_c^2}{6}f_c^{-2}(\phi)e^{-4\alpha+2\sigma_c},\label{Friedmann eq 2}\\
\ddot{\sigma}_b+3\dot{\alpha}\dot{\sigma}_b&=\frac{p_a^2}{3}f_a^{-2}(\phi)e^{-4\alpha-2\sigma_b-2\sigma_c}-\frac{2p_b^2}{3}f_b^{-2}(\phi)e^{-4\alpha+2\sigma_b}+\frac{p_c^2}{3}f_c^{-2}(\phi)e^{-4\alpha+2\sigma_c},\label{Friedmann eq 3}\\
\ddot{\sigma}_c+3\dot{\alpha}\dot{\sigma}_c&=\frac{p_a^2}{3}f_a^{-2}(\phi)e^{-4\alpha-2\sigma_b-2\sigma_c}+\frac{p_b^2}{3}f_b^{-2}(\phi)e^{-4\alpha+2\sigma_b}-\frac{2p_c^2}{3}f_c^{-2}(\phi)e^{-4\alpha+2\sigma_c}.\label{Friedmann eq 4}
\end{align}
In addition, the field equation for the scalar field \eqref{general scalar field eq} takes the form
\begin{align}
\ddot{\phi}+3\dot{\alpha}\dot{\phi}+V'(\phi)-p_a^2 f_a^{-3}(\phi)f_a'(\phi)e^{-4\alpha-2\sigma_b-2\sigma_c}-p_b^2 f_b^{-3}(\phi)f_b'(\phi)e^{-4\alpha+2\sigma_b}-p_c^2 f_c^{-3}(\phi)f_c'(\phi)e^{-4\alpha+2\sigma_c}=0,\label{scalar field eq}
\end{align}
with the help of the solutions shown in Eq. \eqref{vector fields eq}. It is straightforward to verify that Eqs. \eqref{Friedmann eq 1}, \eqref{Friedmann eq 2}, \eqref{Friedmann eq 3}, \eqref{Friedmann eq 4}, and \eqref{scalar field eq} reduce to the field equations derived in Ref. \cite{Kanno:2010nr} if $p_b=p_c=0$ along with $\sigma_b=\sigma_c=\sigma$.
%%%%%%%%%%%%%%%
\section{Anisotropic power-law solutions} \label{sec3}
In order to seek power-law inflationary solutions \cite{Abbott:1984fp,Lucchin:1984yf}, the potential and the coupling functions are chosen to take the following forms \cite{Kanno:2010nr},
\begin{equation}
\begin{aligned}
V(\phi)=V_0 e^{\lambda\phi},\quad f_a(\phi)=f_{a0} e^{\rho_a \phi},\quad f_b(\phi)=f_{b0} e^{\rho_b \phi},\quad f_c(\phi)=f_{c0} e^{\rho_c \phi},
\end{aligned}
\end{equation}
where $V_0$, $f_{a0}$, $f_{b0}$, $f_{c0}$, $\lambda$, $\rho_a$, $\rho_b$, and $\rho_c$ are constants. 
We investigate the power-law solutions in this model by considering the following ansatz \cite{Kanno:2010nr}, 
\begin{equation}
\begin{aligned}
\alpha=\zeta\log t,\quad \sigma_b=\eta_b \log t,\quad \sigma_c=\eta_c\log t,\quad \phi=\xi\log t+\phi_0.
\end{aligned}
\end{equation}
Consequently, the metric \eqref{BI metric} becomes
\begin{align}
ds^2=-dt^2+t^{2\zeta-2\eta_b-2\eta_c}dx^2+t^{2\zeta+2\eta_b}dy^2+t^{2\zeta+2\eta_c}dz^2, \label{power-law BI metric}
\end{align}
which is a power-law metric. The corresponding spatial anisotropies \eqref{anysotropies along axes} are then given by
\begin{align}
X_b=\frac{\eta_b}{\zeta},\quad X_c=\frac{\eta_c}{\zeta},\quad X_a=\frac{\eta_a}{\zeta},
\end{align}
with $\eta_a\equiv-\eta_b-\eta_c$. Apparently, the metric \eqref{power-law BI metric} reduces to the power-law rotationally symmetric BI (rsBI) metric if any two of $\eta_a$, $\eta_b$, and $\eta_c$ are equal, and reduces to the power-law FLRW metric if $\eta_a=\eta_b=\eta_c=0$. 

For convenience, we introduce the additional variables,
\begin{equation}
\begin{aligned}
u=V_0 e^{\lambda\phi_0},\quad \omega_a=p_a^2 f_{a0}^{-2}e^{-2\rho_a\phi_0},\quad \omega_b=p_b^2 f_{b0}^{-2}e^{-2\rho_b\phi_0},\quad \omega_c=p_c^2 f_{c0}^{-2}e^{-2\rho_c\phi_0},
\end{aligned}
\end{equation}
which allow us to rewrite the field equations \eqref{Friedmann eq 1}, \eqref{Friedmann eq 2}, \eqref{Friedmann eq 3}, \eqref{Friedmann eq 4}, and \eqref{scalar field eq} as
\begin{equation}\label{general power-law system of eqs}
\begin{aligned}
\frac{-3\zeta^2+\eta_b^2+\eta_c^2+\eta_b \eta_c+\xi^2/2}{t^2}+u t^{\lambda\xi}+\frac{\omega_a}{2t^{2(\rho_a\xi+2\zeta+\eta_b+\eta_c)}}+\frac{\omega_b}{2t^{2(\rho_b\xi+2\zeta-\eta_b)}}+\frac{\omega_c}{2t^{2(\rho_c\xi+2\zeta-\eta_c)}}&=0,\\
\frac{-\zeta+3\zeta^2}{t^2}-u t^{\lambda\xi}-\frac{\omega_a}{6t^{2(\rho_a\xi+2\zeta+\eta_b+\eta_c)}}-\frac{\omega_b}{6t^{2(\rho_b\xi+2\zeta-\eta_b)}}-\frac{\omega_c}{6t^{2(\rho_c\xi+2\zeta-\eta_c)}}&=0,\\
\frac{-\eta_b+3\zeta\eta_b}{t^2}-\frac{\omega_a}{3t^{2(\rho_a\xi+2\zeta+\eta_b+\eta_c)}}+\frac{2\omega_b}{3t^{2(\rho_b\xi+2\zeta-\eta_b)}}-\frac{\omega_c}{3t^{2(\rho_c\xi+2\zeta-\eta_c)}}&=0,\\
\frac{-\eta_c+3\zeta\eta_c}{t^2}-\frac{\omega_a}{3t^{2(\rho_a\xi+2\zeta+\eta_b+\eta_c)}}-\frac{\omega_b}{3t^{2(\rho_b\xi+2\zeta-\eta_b)}}+\frac{2\omega_c}{3t^{2(\rho_c\xi+2\zeta-\eta_c)}}&=0,\\
\frac{-\xi+3\zeta\xi}{t^2}+\lambda u t^{\lambda\xi}-\frac{\rho_a\omega_a}{t^{2(\rho_a\xi+2\zeta+\eta_b+\eta_c)}}-\frac{\rho_b\omega_b}{t^{2(\rho_b\xi+2\zeta-\eta_b)}}-\frac{\rho_c\omega_c}{t^{2(\rho_c\xi+2\zeta-\eta_c)}}&=0.
\end{aligned}
\end{equation}
Remarkably, the field equations are no longer ordinary differential equations but algebraic ones. In the following, by imposing suitable constraints, we will find four types of solutions that are classified according to the number of non-vanishing vector fields.
%%%%%%%%%%%%%%%
\subsection{Solution type 0}
In the case where all the vector fields vanish, i.e., $p_a=p_b=p_c=0$, the system \eqref{general power-law system of eqs} becomes
\begin{equation}\label{power-law system of eqs vec0}
\begin{aligned}
-\lambda\xi&=2,\\
3\zeta^2-\eta_b^2-\eta_c^2-\eta_b\eta_c-\frac{\xi^2}{2}-u&=0,\\
\zeta-3\zeta^2+u&=0,\\
\eta_b-3\zeta\eta_b&=0,\\
\eta_c-3\zeta\eta_c&=0,\\
\xi-3\zeta\xi-\lambda u&=0,
\end{aligned}
\end{equation}
where the first equation is required to ensure that all terms in each equation of the system \eqref{general power-law system of eqs} have the same time dependence. The system \eqref{power-law system of eqs vec0} has the solution
\begin{equation}\label{power-law solution type 0}
\begin{aligned}
\zeta=\frac{2}{\lambda^2},\quad \eta_a=\eta_b=\eta_c=0,\quad \xi=-\frac{2}{\lambda},\quad u=\frac{2(6-\lambda^2)}{\lambda^4}.
\end{aligned}
\end{equation}
Since Eq. \eqref{power-law solution type 0} corresponds to the solution with no non-vanishing vector field, it is convenient to refer to it as the solution type $0$. We notice that $\eta_a=\eta_b=\eta_c=0$ for this solution. This means that it corresponds to a power-law FLRW metric,
\begin{align}
ds^2=-dt^2+t^{4/\lambda^2}(dx^2+dy^2+dz^2),
\end{align}
where the scale factors are given by $a(t)=b(t)=c(t)=t^{2/\lambda^2}$. This type of solution can be found in Ref. \cite{Kanno:2010nr} as well as in the original model of power-law inflation \cite{Lucchin:1984yf}.
%%%%%%%%%%%%%%%%
\subsection{Solutions type I}
In the case where $A_\mu$ is the only non-vanishing vector field, i.e., $p_a\neq 0$ and $p_b=p_c=0$, the system \eqref{general power-law system of eqs} becomes
\begin{equation}\label{power-law system of eqs vec1x}
\begin{aligned}
-\lambda\xi=2,\quad\rho_a\xi+2\zeta+\eta_b+\eta_c&=1,\\
3\zeta^2-\eta_b^2-\eta_c^2-\eta_b\eta_c-\frac{\xi^2}{2}-u-\frac{\omega_a}{2}&=0,\\
\zeta-3\zeta^2+u+\frac{\omega_a}{6}&=0,\\
\eta_b-3\zeta\eta_b+\frac{\omega_a}{3}&=0,\\
\eta_c-3\zeta\eta_c+\frac{\omega_a}{3}&=0,\\
\xi-3\zeta\xi-\lambda u+\rho_a \omega_a&=0,
\end{aligned}
\end{equation} 
where the first two equations are required to ensure that all terms in each equation of the system \eqref{general power-law system of eqs} have the same time dependence. The system \eqref{power-law system of eqs vec1x} has the following solution,
\begin{equation}\label{power-law solution type Ia}
\begin{aligned}
\zeta&=\frac{8+\lambda^2+8\lambda\rho_a+12\rho_a^2}{6\lambda(\lambda+2\rho_a)},\quad \eta_a=\frac{2(4-\lambda^2-2\lambda\rho_a)}{3\lambda(\lambda+2\rho_a)},\quad \eta_b=\eta_c=\frac{-4+\lambda^2+2\lambda\rho_a}{3\lambda(\lambda+2\rho_a)},\quad \xi=-\frac{2}{\lambda},\\
u&=\frac{(2+\lambda\rho_a+2\rho_a^2)(8-\lambda^2+4\lambda\rho_a+12\rho_a^2)}{2\lambda^2(\lambda+2\rho_a)^2},\quad
\omega_a=\frac{(-4+\lambda^2+2\lambda\rho_a)(8-\lambda^2+4\lambda\rho_a+12\rho_a^2)}{2\lambda^2(\lambda+2\rho_a)^2}.
\end{aligned}
\end{equation}
Since Eq. \eqref{power-law solution type Ia} is the solution where $A_\mu$ is the only non-vanishing vector field ($p_a\neq 0$ and $p_b=p_c=0$), we refer to it as the solution type $\text{I}_a$. We also notice that $\eta_b=\eta_c$ for this solution, so it corresponds to the power-law $x$-rsBI metric mentioned above. In fact, this solution is nothing but the anisotropic solution found in Ref. \cite{Kanno:2010nr}. Similarly, in the case where $B_\mu$ is the only non-vanishing vector field, i.e., $p_b\neq 0$ and $p_a=p_c=0$, it is straightforward to obtain the solution type $\text{I}_b$, which corresponds to a power-law $y$-rsBI metric, by interchanging indices $a\leftrightarrow b$ in Eq. \eqref{power-law solution type Ia}. Finally, in the case where $C_\mu$ is the only non-vanishing vector field, i.e., $p_c\neq 0$ and $p_a=p_b=0$, we obtain the solution type $\text{I}_c$, which corresponds to a power-law $z$-rsBI metric, by interchanging indices $a\leftrightarrow c$ in Eq. \eqref{power-law solution type Ia}.
%%%%%%%%%%%%%%%%%%
\subsection{Solutions type II}
In the case where $B_\mu$ and $C_\mu$ are the only two non-vanishing vector fields, i.e., $p_a=0$, $p_b\neq 0$, and $p_c\neq 0$, the system \eqref{general power-law system of eqs} becomes
\begin{equation}\label{power-law system of eqs vec2yz}
\begin{aligned}
-\lambda\xi=2,\quad
\rho_b\xi+2\zeta-\eta_b=1,\quad
\rho_c\xi+2\zeta-\eta_c&=1,\\
3\zeta^2-\eta_b^2-\eta_c^2-\eta_b \eta_c-\frac{\xi^2}{2}-u-\frac{\omega_b}{2}-\frac{\omega_c}{2}&=0,\\
\zeta-3\zeta^2+u+\frac{\omega_b}{6}+\frac{\omega_c}{6}&=0,\\
\eta_b-3\zeta\eta_b-\frac{2\omega_b}{3}+\frac{\omega_c}{3}&=0,\\
\eta_c-3\zeta\eta_c+\frac{\omega_b}{3}-\frac{2\omega_c}{3}&=0,\\
\xi-3\zeta\xi-\lambda u+\rho_b\omega_b+\rho_c\omega_c&=0,
\end{aligned}
\end{equation}
where the first three equations are required to ensure that all terms in each equation of the system \eqref{general power-law system of eqs} have the same time dependence. The system \eqref{power-law system of eqs vec2yz} has the following solution,
\begin{equation}\label{power-law solution type IIbc}
\begin{aligned}
\zeta&=\frac{2+\lambda^2+4\lambda(\rho_b+\rho_c)+4(\rho_b^2+\rho_c^2+\rho_b\rho_c)}{3\lambda\left[\lambda+2(\rho_b+\rho_c)\right]},\quad
\eta_a=\frac{2\left[-4+\lambda^2+\lambda(\rho_b+\rho_c)-2(\rho_b^2+\rho_c^2-2\rho_b\rho_c)\right]}{3\lambda\left[\lambda+2(\rho_b+\rho_c)\right]},\\
\eta_b&=\frac{4-\lambda^2-2\lambda(2\rho_b-\rho_c)-4(\rho_b^2-2\rho_c^2+\rho_b\rho_c)}{3\lambda\left[\lambda+2(\rho_b+\rho_c)\right]},\quad
\eta_c=\frac{4-\lambda^2-2\lambda(2\rho_c-\rho_b)-4(\rho_c^2-2\rho_b^2+\rho_b\rho_c)}{3\lambda\left[\lambda+2(\rho_b+\rho_c)\right]},\\
\xi&=-2/\lambda,\quad u=\frac{2\left[2+\lambda(\rho_b+\rho_c)+2(\rho_b^2+\rho_c^2)\right]\left[1+\lambda(\rho_b+\rho_c)+2(\rho_b^2+\rho_c^2+\rho_b \rho_c)\right]}{\lambda^2\left[\lambda+2(\rho_b+\rho_c)\right]^2},\\
\omega_b&=\frac{2\left[-4+\lambda^2+2\lambda\rho_b-4\rho_c(\rho_c-\rho_b)\right]\left[1+\lambda(\rho_b+\rho_c)+2(\rho_b^2+\rho_c^2+\rho_b\rho_c)\right]}{\lambda^2\left[\lambda+2(\rho_b+\rho_c)\right]^2},\\
\omega_c&=\frac{2\left[-4+\lambda^2+2\lambda\rho_c-4\rho_b(\rho_b-\rho_c)\right]\left[1+\lambda(\rho_b+\rho_c)+2(\rho_b^2+\rho_c^2+\rho_b\rho_c)\right]}{\lambda^2\left[\lambda+2(\rho_b+\rho_c)\right]^2}.
\end{aligned}
\end{equation}
Since Eq. \eqref{power-law solution type IIbc} is the solution where $B_\mu$ and $C_\mu$ are the only two non-vanishing vector fields, we refer to it as the solution type $\text{II}_{bc}$. In general, $\eta_a$, $\eta_b$, and $\eta_c$ are different, so the solution type $\text{II}_{bc}$ corresponds to a general power-law BI metric with three different scale factors. Note that if $\rho_b=\rho_c$, then $\eta_b=\eta_c$ and the metric becomes a power-law $x$-rsBI metric. Similarly, in the case where $A_\mu$ and $C_\mu$ are the only two non-vanishing vector fields, i.e., $p_a\neq 0$, $p_b= 0$, and $p_c\neq 0$, we are able to obtain the corresponding solution type $\text{II}_{ac}$ by interchanging indices $a\leftrightarrow b$ in Eq. \eqref{power-law solution type IIbc}. Finally, in the case where $A_\mu$ and $B_\mu$ are the only two non-vanishing vector fields, i.e., $p_a\neq 0$, $p_b\neq 0$, and $p_c= 0$, we obtain the corresponding solution type $\text{II}_{ab}$ by interchanging indices $a\leftrightarrow c$ in Eq. \eqref{power-law solution type IIbc}.
%%%%%%%%%%%%%%%%%%%%
\subsection{Solution type III}
In the case where $A_\mu$, $B_\mu$, and $C_\mu$ are all non-vanishing vector fields, i.e., $p_a\neq 0$, $p_b\neq 0$, and $p_c\neq 0$, the system \eqref{general power-law system of eqs} becomes
\begin{equation}\label{power-law system of eqs vec3}
\begin{aligned}
-\lambda\xi=2,\quad
\rho_a\xi+2\zeta+\eta_b+\eta_c =1,\quad
 \rho_b\xi+2\zeta-\eta_b=1,\quad
 \rho_c\xi+2\zeta-\eta_c& =1,\\
-3\zeta^2+\eta_b^2+\eta_c^2+\eta_b \eta_c+\frac{\xi^2}{2}+u+\frac{\omega_a}{2}+\frac{\omega_b}{2}+\frac{\omega_c}{2}&=0,\\
-\zeta+3\zeta^2-u-\frac{\omega_a}{6}-\frac{\omega_b}{6}-\frac{\omega_c}{6}&=0,\\
-\eta_b+3\zeta\eta_b-\frac{\omega_a}{3}+\frac{2\omega_b}{3}-\frac{\omega_c}{3}&=0,\\
-\eta_c+3\zeta\eta_c-\frac{\omega_a}{3}-\frac{\omega_b}{3}+\frac{2\omega_c}{3}&=0,\\
-\xi+3\zeta\xi+\lambda u-\rho_a\omega_a-\rho_b\omega_b-\rho_c\omega_c&=0,
\end{aligned}
\end{equation}
where the first four equations are required to ensure that all terms in each equation of the system \eqref{general power-law system of eqs} have the same time dependence. The system \eqref{power-law system of eqs vec3} has the following solution,
\begin{equation} \label{power-law solution vec3}
\begin{aligned}
\zeta&=\frac{3\lambda+2\Sigma_\rho}{6\lambda},\quad
\eta_i=\frac{2(\Sigma_\rho-3\rho_i)}{3\lambda},\quad
\xi=-\frac{2}{\lambda},\\
u&=\frac{2+\lambda\Sigma_\rho+2\Sigma_{\rho^2}}{2\lambda^2},\quad
\omega_i=\frac{-4+\lambda^2+2\lambda\rho_i-4\left(\Sigma_{\rho^2}-\rho_i\Sigma_\rho\right)}{2\lambda^2},\quad\text{for }i=a,b,c.
\end{aligned}
\end{equation}
Here, we have defined
\begin{align}\label{definition of Sigma}
\Sigma_\rho\equiv\rho_a+\rho_b+\rho_c,\quad\Sigma_{\rho^2}\equiv\rho_a^2+\rho_b^2+\rho_c^2.
\end{align}
Since \eqref{power-law solution vec3} is the solution where all three vector fields $A_\mu$, $B_\mu$, and $C_\mu$ are non-vanishing, we refer to it as the solution type III. In general, $\eta_a$, $\eta_b$, and $\eta_c$ are different, so the solution type $\text{III}$ corresponds to a general power-law BI metric with three different scale factors. If two of the three coupling constants are equal, say $\rho_b$ and $\rho_c$, then $\eta_b=\eta_c$ and the metric becomes a power-law $x$-rsBI metric. If all three coupling constants are equal, i.e., $\rho_a=\rho_b=\rho_c$, then $\eta_a=\eta_b=\eta_c$ and the metric becomes a power-law FLRW metric.

For convenience, we summarize all the solutions obtained above in Table \eqref{tab: power-law solutions}.
\renewcommand{\arraystretch}{1.3}
\begin{table}[H]
\begin{ruledtabular}
\begin{tabular}{cccc}
\textbf{Solution} & \multirow[c]{2}{*}{\textbf{Specific formula}} & \textbf{Non-vanishing} & \textbf{Corresponding}\\
\textbf{type} & & \textbf{vector field(s)} & \textbf{metric}\\
\hline
0 & \eqref{power-law solution type 0} & None & FLRW\\
\hline
$\text{I}_a$ & \eqref{power-law solution type Ia} & $A_\mu$ & $x$-rsBI\\
\hline
$\text{I}_b$ & \eqref{power-law solution type Ia} with $a\leftrightarrow b$ & $B_\mu$ & $y$-rsBI\\
\hline
$\text{I}_c$ & \eqref{power-law solution type Ia} with $a\leftrightarrow c$ & $C_\mu$ & $z$-rsBI\\
\hline
$\text{II}_{bc}$ & \eqref{power-law solution type IIbc} & $B_\mu$ and $C_\mu$ & general BI\\
\hline
$\text{II}_{ac}$ & \eqref{power-law solution type IIbc} with $a\leftrightarrow b$ & $A_\mu$ and $C_\mu$ & general BI\\
\hline
$\text{II}_{ab}$ & \eqref{power-law solution type IIbc} with $a\leftrightarrow c$ & $A_\mu$ and $B_\mu$ & general BI\\
\hline
III & \eqref{power-law solution vec3} & $A_\mu$, $B_\mu$, and $C_\mu$ & general BI\\
\end{tabular}
\end{ruledtabular}
\caption{\label{tab: power-law solutions} List of the power-law solutions.}
\end{table}
%%%%%%%%%%%%%%%%%%
\section{Stability analysis}\label{sec4}
\subsection{Dynamical system and fixed points}
For the stability analysis of the derived solutions, we would like to introduce the following dynamical variables \cite{Kanno:2010nr}, 
\begin{equation}\label{dynamical variables} 
\begin{aligned}
X_b&= \frac{\dot{\sigma}_b}{\dot{\alpha}},\quad X_c=\frac{\dot{\sigma}_c}{\dot{\alpha}},\quad X_a= -X_b-X_c=\frac{-\dot{\sigma}_b-\dot{\sigma}_c}{\dot{\alpha}},\quad Y=\frac{\dot{\phi}}{\dot{\alpha}},\\
Z_a&= \frac{p_a f_a^{-1}}{\dot{\alpha}}e^{-2\alpha-\sigma_b-\sigma_c},\quad Z_b= \frac{p_b f_b^{-1}}{\dot{\alpha}}e^{-2\alpha+\sigma_b},\quad Z_c = \frac{p_c f_c^{-1}}{\dot{\alpha}}e^{-2\alpha+\sigma_c},
\end{aligned}
\end{equation}
where $X_a$ is an auxiliary variable. Note that the variables $X_a$, $X_b$, and $X_c$, which measure the anisotropies along $x$-, $y$-, and $z$-axes, have been defined in \eqref{anysotropies along axes}. Thanks to \eqref{dynamical variables} and the field equations \eqref{Friedmann eq 1}, \eqref{Friedmann eq 2}, \eqref{Friedmann eq 3}, \eqref{Friedmann eq 4}, and \eqref{scalar field eq}, we can obtain a system of autonomous equations of the dynamical variables. Next, we seek fixed points of the dynamical system by solving the set of equations,
\begin{align}\label{fixed point system of eqs}
\frac{dX_b}{d\alpha}=\frac{dX_c}{d\alpha}=\frac{dY}{d\alpha}=\frac{dZ_a}{d\alpha}=\frac{dZ_b}{d\alpha}=\frac{dZ_c}{d\alpha}=0.
\end{align}
The detailed derivations of the system and the corresponding fixed points will be presented in Appendix \ref{Appendix 1}. In the following, we only show the main results.

For $Z_a=Z_b=Z_c=0$, solving Eq. \eqref{fixed point system of eqs} gives us an isotropic fixed point given by
\begin{align}
X_a=X_b=X_c=0,\quad Y=-\lambda,\quad Z_a=Z_b=Z_c=0.\label{fixed point type 0}
\end{align}
As one can easily check, the fixed point \eqref{fixed point type 0} corresponds to the solutions type 0 shown in Eq. \eqref{power-law solution type 0}. Therefore, it is convenient to refer to it as the fixed point type 0. 

For $Z_a\neq 0$ and $Z_b=Z_c=0$, solving \eqref{fixed point system of eqs} gives us an anisotropic fixed point given by
\begin{equation}\label{fixed point type Ia}
\begin{aligned}
X_a&=\frac{4(4-\lambda^2-2\lambda\rho_a)}{8+\lambda^2+8\lambda\rho_a+12\rho_a^2},\quad
X_b=X_c=\frac{2(-4+\lambda^2+2\lambda\rho_a)}{8+\lambda^2+8\lambda\rho_a+12\rho_a^2},\quad
Y=\frac{-12(\lambda+2\rho_a)}{8+\lambda^2+8\lambda\rho_a+12\rho_a^2},\\
\left(Z_a\right)^2&=\frac{18(-4+\lambda^2+2\lambda\rho_a)(8-\lambda^2+4\lambda\rho_a+12\rho_a^2)}{(8+\lambda^2+8\lambda\rho_a+12\rho_a^2)^2},\quad
Z_b=Z_c=0,
\end{aligned}
\end{equation}
whose existence region is the following inequality,
\begin{align}
(-4+\lambda^2+2\lambda\rho_a)(8-\lambda^2+4\lambda\rho_a+12\rho_a^2)>0,\label{fixed point type Ia ex cond}
\end{align}
to ensure that $(Z_a)^2>0$. One can verify that the fixed point \eqref{fixed point type Ia} corresponds to the solution type $\text{I}_a$ \eqref{power-law solution type Ia}. Therefore, it is convenient to refer to the fixed point \eqref{fixed point type Ia} as the fixed point type $\text{I}_a$. Similarly, for $Z_b\neq 0$ and $Z_a=Z_c=0$, we have the fixed point type $\text{I}_b$ and its existence region obtained by interchanging indices $a\leftrightarrow b$ in \eqref{fixed point type Ia} and \eqref{fixed point type Ia ex cond}. Finally, we have the fixed point type $\text{I}_c$ and its existence region for $Z_c\neq 0$ and $Z_a=Z_b=0$ by interchanging indices $a\leftrightarrow c$ in \eqref{fixed point type Ia} and \eqref{fixed point type Ia ex cond}.

For $Z_a=0$, $Z_b \neq 0$, and $Z_c\neq 0$, solving Eq. \eqref{fixed point system of eqs} gives us an anisotropic fixed point given by
\begin{equation}\label{fixed point type IIbc}
\begin{aligned}
X_a&=\frac{-8+2\lambda^2+2\lambda(\rho_b+\rho_c)-4(\rho_b^2+\rho_c^2-2\rho_b\rho_c)}{2+\lambda^2+4\left[\lambda(\rho_b+\rho_c)+\rho_b^2+\rho_c^2+\rho_b\rho_c\right]},\quad
X_b=\frac{4-\lambda^2-2\lambda(2\rho_b-\rho_c)-4(\rho_b^2-2\rho_c^2+\rho_b\rho_c)}{2+\lambda^2+4\left[\lambda(\rho_b+\rho_c)+\rho_b^2+\rho_c^2+\rho_b\rho_c\right]},\\
X_c&=\frac{4-\lambda^2-2\lambda(2\rho_c-\rho_b)-4(\rho_c^2-2\rho_b^2+\rho_b\rho_c)}{2+\lambda^2+4\left[\lambda(\rho_b+\rho_c)+\rho_b^2+\rho_c^2+\rho_b\rho_c\right]},\quad
Y=\frac{-6\left[\lambda+2(\rho_b+\rho_c)\right]}{2+\lambda^2+4\left[\lambda(\rho_b+\rho_c)+\rho_b^2+\rho_c^2+\rho_b\rho_c\right]},\\
Z_a&=0,\quad
\left(Z_b\right)^2=\frac{18\left[-4+\lambda^2+2\lambda\rho_b-4\rho_c(\rho_c-\rho_b)\right]\left[1+\lambda(\rho_b+\rho_c)+2(\rho_b^2+\rho_c^2+\rho_b\rho_c)\right]}{\left[2+\lambda^2+4\lambda(\rho_b+\rho_c)+4(\rho_b^2+\rho_c^2+\rho_b\rho_c)\right]^2},\\
\left(Z_c \right)^2&=\frac{18\left[-4+\lambda^2+2\lambda\rho_c-4\rho_b(\rho_b-\rho_c)\right]\left[1+\lambda(\rho_b+\rho_c)+2(\rho_b^2+\rho_c^2+\rho_b\rho_c)\right]}{\left[2+\lambda^2+4\lambda(\rho_b+\rho_c)+4(\rho_b^2+\rho_c^2+\rho_b\rho_c)\right]^2},
\end{aligned}
\end{equation}
whose the existence region is determined by two inequalities, 
\begin{equation}\label{fixed point type IIbc ex cond}
\begin{aligned}
\left[-4+\lambda^2+2\lambda\rho_b-4\rho_c(\rho_c-\rho_b)\right]\left[1+\lambda(\rho_b+\rho_c)+2(\rho_b^2+\rho_c^2+\rho_b\rho_c)\right]&>0,\\
\left[-4+\lambda^2+2\lambda\rho_c-4\rho_b(\rho_b-\rho_c)\right]\left[1+\lambda(\rho_b+\rho_c)+2(\rho_b^2+\rho_c^2+\rho_b\rho_c)\right]&>0.
\end{aligned}
\end{equation}
As a result, these constraints are due to the positivity of $(Z_b)^2$ and $(Z_c)^2$. Since the fixed point \eqref{fixed point type IIbc} is equivalent to the solution type $\text{II}_{bc}$ \eqref{power-law solution type IIbc}, we will refer to it as the fixed point type $\text{II}_{bc}$ for convenience. Similarly, for $Z_b=0$ and $Z_a,Z_c\neq 0$, we have the fixed point type $\text{II}_{ac}$ and its existence region by interchanging indices $a\leftrightarrow b$ in \eqref{fixed point type IIbc} and \eqref{fixed point type IIbc ex cond}. Similarly, for $Z_c=0$ and $Z_a,Z_b\neq 0$, we have the fixed point type $\text{II}_{ab}$, obtained by interchanging indices $a\leftrightarrow c$ in \eqref{fixed point type IIbc} and \eqref{fixed point type IIbc ex cond}.

For $Z_i\neq 0$ for $i=a,b,c$, solving \eqref{fixed point system of eqs} gives us the fixed point type III, which corresponds to the solution type III, as
\begin{align}\label{fixed point type III}
X_i&=\frac{4(\Sigma_\rho-3\rho_i)}{3\lambda+2\Sigma_\rho},\quad
Y=\frac{-12}{3\lambda+2\Sigma_\rho},\quad
\left(Z_i \right)^2=\frac{18\left[-4+\lambda^2+2\lambda\rho_i-4\left(\Sigma_{\rho^2}-\rho_i\Sigma_\rho\right)\right]}{\left(3\lambda+2\Sigma_\rho\right)^2},
\end{align}
of which the existence region is
\begin{equation}\label{fixed point type III ex cond}
\begin{aligned}
-4+\lambda^2+2\lambda\rho_i-4\left(\Sigma_{\rho^2}-\rho_i\Sigma_\rho\right)>0,\\
\end{aligned}
\end{equation}
to ensure the positivity of $(Z_i)^2$. Note that $\Sigma_\rho$ and $\Sigma_{\rho^2}$ have been defined in \eqref{definition of Sigma}.
%%%%%%%%%%%%%%%
\subsection{Stability regions}
We have found the fixed points of the system of autonomous equations \eqref{autonomous eq Xy}-\eqref{autonomous eq Zz} along with their existence regions. Now we would like to investigate whether the fixed points are stable. Particularly, we will find the stability region of each fixed point (the condition under which the fixed point is stable), which is described by a set of constraints on parameters $\lambda$, $\rho_a$, $\rho_b$ and $\rho_c$. The standard procedure is to perturb the system \eqref{autonomous eq Xy}-\eqref{autonomous eq Zz} to first order to obtain 
\begin{align}\label{perturbed dynamical system}
\frac{d\mathbf{u}}{d\alpha} = M \mathbf{u}.
\end{align}
where $\mathbf{u}\equiv(\delta X_b,\delta X_c,\delta Y,\delta Z_a,\delta Z_b,\delta Z_c)^T$ and $M$ is a $6\times 6$ matrix, whose entries will be given in Appendix \ref{RH criterion}. The next step is to substitute each fixed point into the eigenvalue equation,
\begin{align}\label{eigenvalue eq}
\det(M-sI_6)=0,
\end{align}
where $I_6$ is the $6\times 6$ identity matrix. The stability region is then determined by the conditions on parameters that ensure that all the roots of Eq. \eqref{eigenvalue eq} have negative real parts. For the fixed point type 0, the task is simple since the corresponding matrix $M$ is diagonal. For the other fixed points, especially for the fixed point type III, the eigenvalue equation becomes very complicated to solve analytically. In principle, one may choose several specific sets of parameter values and solve the eigenvalue equation numerically. However, this approach would not provide a systematic picture of the stability of the fixed points. Therefore, we instead employ the Routh-Hurwitz criterion - a helpful technique in stability analysis \cite{Merkin,Nise}. The power of the Routh-Hurwitz criterion is that it tells us whether all the roots have negative real parts without explicitly solving the equation. The detailed derivation of the stability regions of the fixed points will be presented in Appendix \ref{RH criterion} for convenience. In the following, we will only describe the main results. For simplicity, we only consider the case where $\lambda$, $\rho_a$, $\rho_b$, and $\rho_c$ are positive from now on.

\subsubsection{Fixed point type 0}
\begin{figure}[H]
\centering
\includegraphics[width=0.3\textwidth]{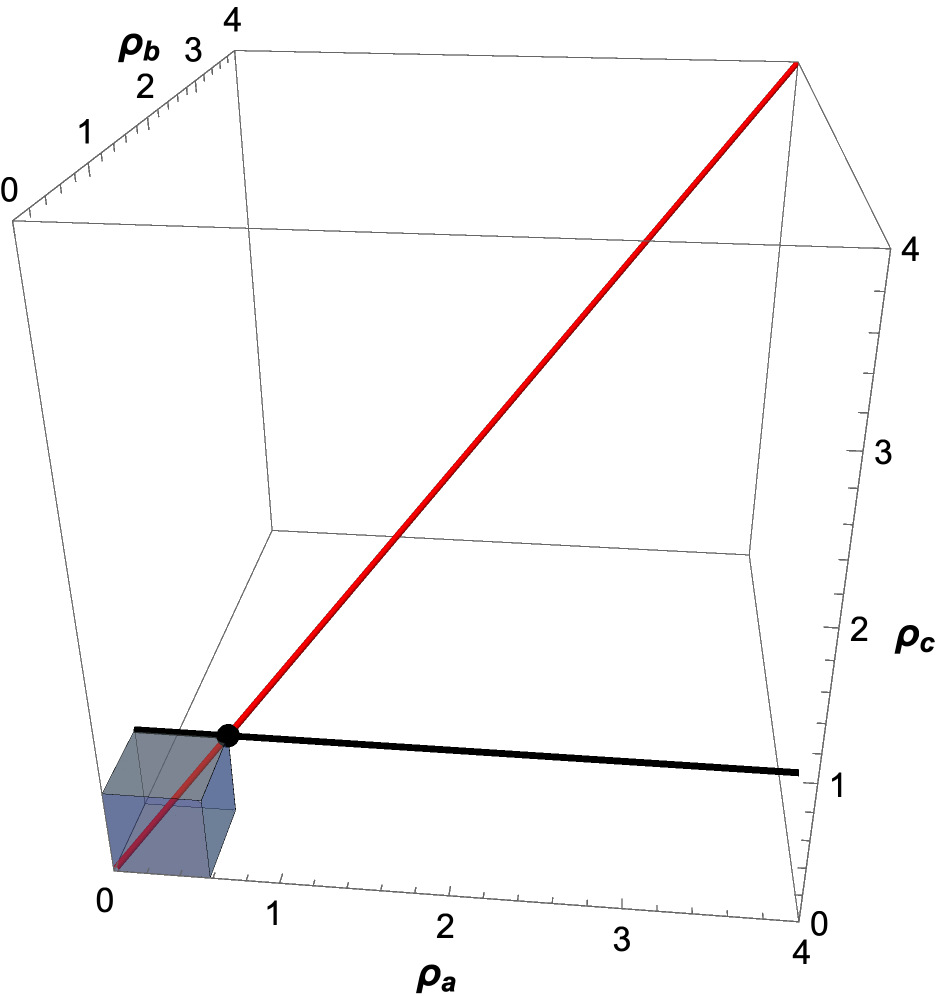}
\caption{Stability region  of the fixed point type 0 for $\lambda=1.5$ displayed as the dark blue cube. The red line corresponds to $\rho_a=\rho_b=\rho_c$. The black point is a special point, where $\rho_a=\rho_b=\rho_c=(4-\lambda^2)/(2\lambda)$. The thick black line corresponds to $\rho_b=\rho_c=(4-\lambda^2)/(2\lambda)$.}
\label{fig:Vec0StRegion}
\end{figure}

The stability region of the fixed point type 0 is determined by the following inequalities, 
\begin{align}\label{fixed point type 0 st cond}
\lambda^2 + 2\lambda \rho_i - 4 < 0,\quad\text{for }i = a,b,c.
\end{align}
For a fixed $\lambda$, the stability region is a cube bounded by three planes $\rho_i=(4-\lambda^2)/(2\lambda)$. Qualitatively, in order to make the fixed point type 0 stable, $\rho_i$ parameters have to be sufficiently small. For illustration, the stability region of the fixed point type 0 for a specific value $\lambda=1.5$ is plotted in Fig. \ref{fig:Vec0StRegion}.

%%%%%%%%%%%%%%%%%
\subsubsection{Fixed point type III}
\begin{figure}[H]
\centering
\includegraphics[width=0.3\textwidth]{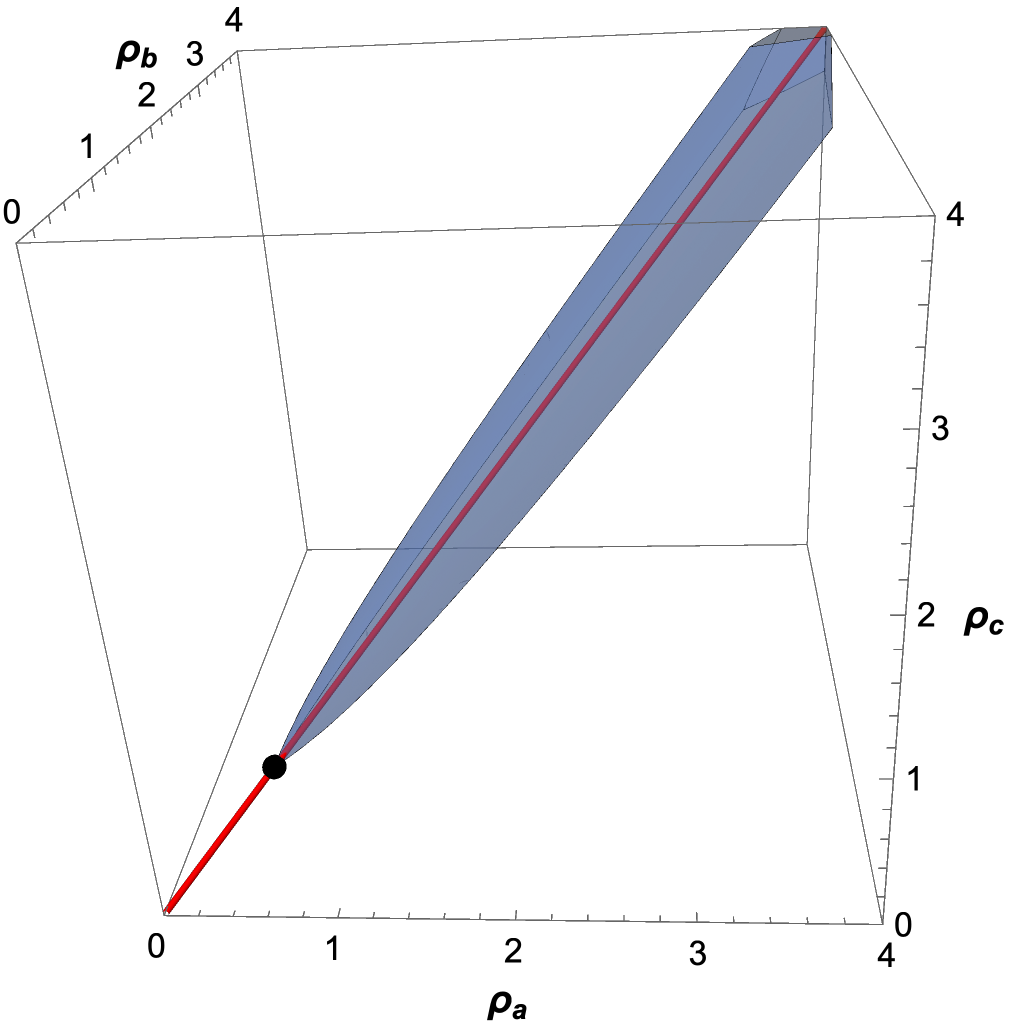}
\caption{Existence and stability region of the fixed point type III for $\lambda=1.5$ displayed as the dark blue region. The red line corresponds to $\rho_a=\rho_b=\rho_c$, while the black point corresponds to $\rho_a=\rho_b=\rho_c=(4-\lambda^2)/(2\lambda)$.}
\label{fig:Vec3StRegion}
\end{figure}

In contrast to the fixed point type 0, which is clearly isotropic, the fixed point type III is generically anisotropic. The existence region of the fixed point type III is determined by three inequalities shown in Eq. \eqref{fixed point type III ex cond} and depicted as the dark blue region  in Fig. \ref{fig:Vec3StRegion} for $\lambda=1.5$. This region is bounded by three surfaces $-4+\lambda^2+2\lambda\rho_i-4\left(\Sigma_{\rho^2}-\rho_i\Sigma_\rho\right)=0$ for $i=a,b,c$. Apparently, for the fixed point type III as well as the other fixed points, the stability region has to be included in the existence region since a fixed point must first exist if we are to examine its stability. Interestingly, we have found that the stability region of the fixed point type III coincides exactly with its existence region determined by Eq. \eqref{fixed point type III ex cond}. This means that the stability region is also depicted as the dark blue region in Fig. \ref{fig:Vec3StRegion} for $\lambda=1.5$. 

It is useful to provide a qualitative description of this stability region. We see that it is a narrow region surrounding the line $\rho_a=\rho_b=\rho_c$, which means that the coupling constants $\rho_i$ have to be sufficiently close to each other in order to make the fixed point type III stable. Additionally, the fact that the region terminates at the black point $\rho_a=\rho_b=\rho_c=(4-\lambda^2)/(2\lambda)$ (note that it is identical to the black point in Fig. \ref{fig:Vec0StRegion}) implies that all the coupling constants $\rho_i$ also have to be sufficiently large. In contrast, if all the coupling constants are smaller than $(4-\lambda^2)/(2\lambda)$, then the fixed point type III cannot be stable, regardless of how close they are.
%%%%%%%%%%%%%%%%%
\subsubsection{Fixed points type II}
\begin{figure}[H]
\centering

\begin{subfigure}{0.3\textwidth}
  \centering
  \includegraphics[width=\textwidth]{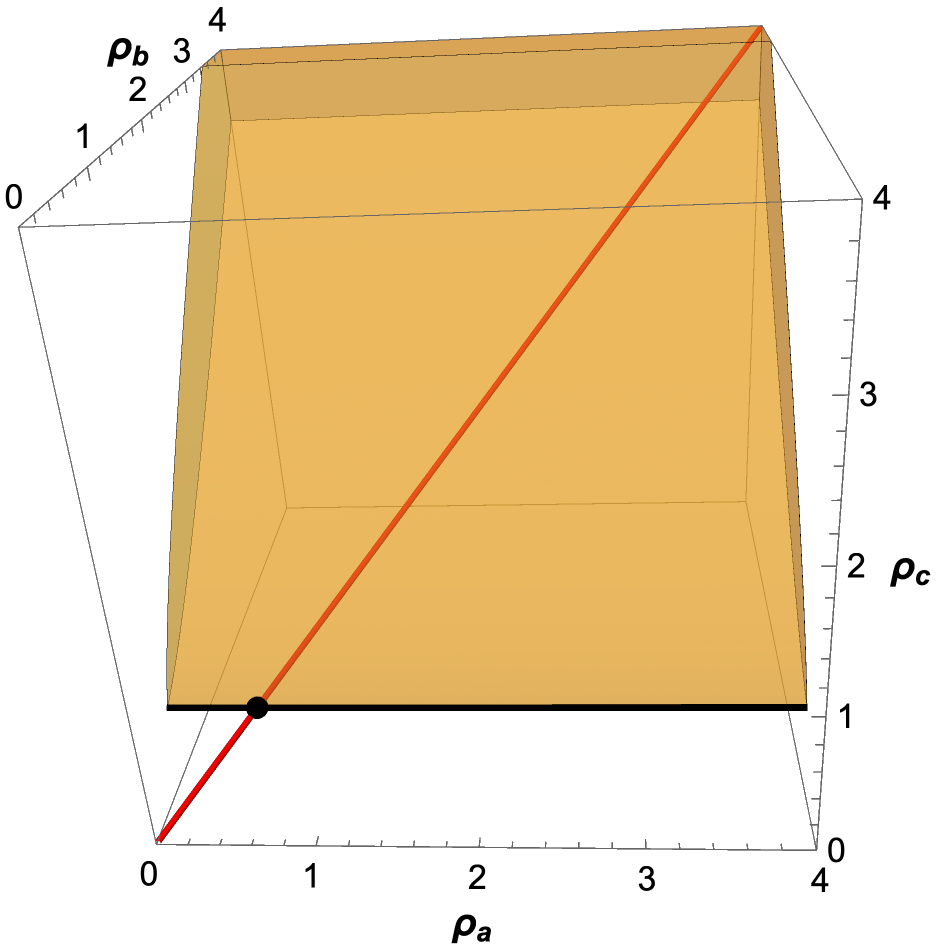}
  \caption{}
  \label{fig:Vec2yzExRegion}
\end{subfigure}
\qquad\qquad\qquad
\begin{subfigure}{0.3\textwidth}
  \centering
  \includegraphics[width=\textwidth]{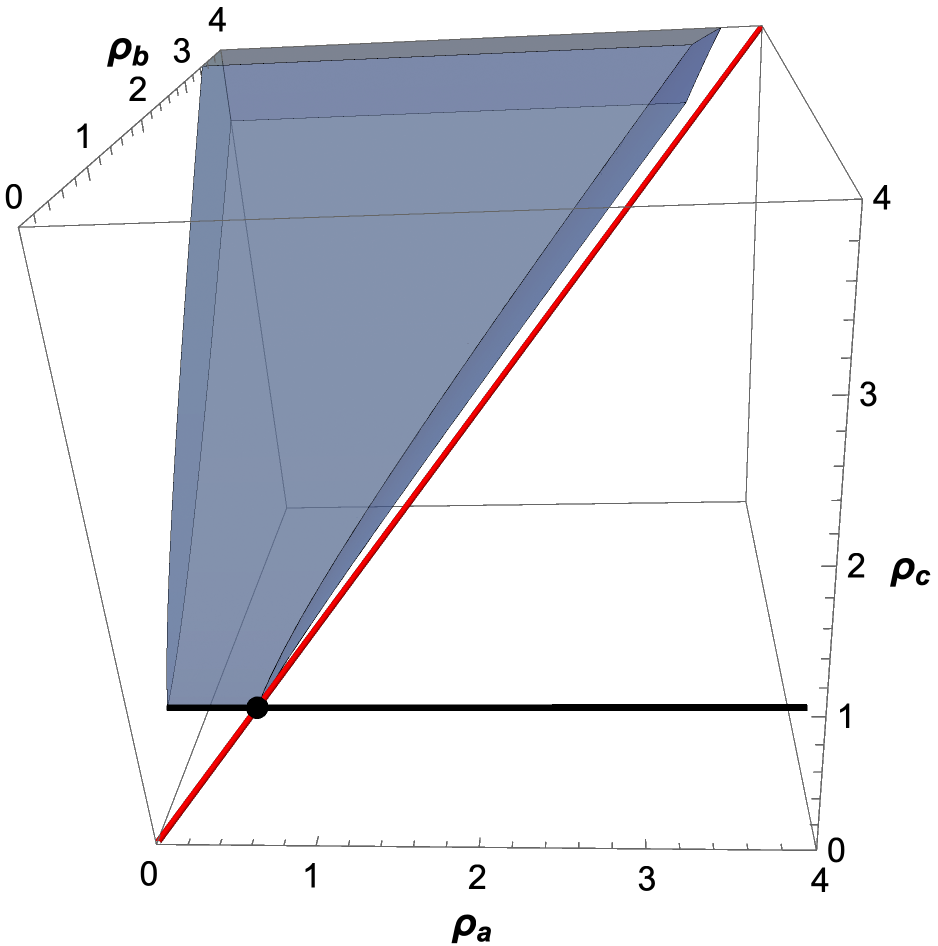}
  \caption{}
  \label{fig:Vec2yzStRegion}
\end{subfigure}

\caption{(left) Existence region colored as orange and (right) stability region colored as dark blue of the fixed point type $\text{II}_{bc}$ for $\lambda=1.5$. The red line corresponds to $\rho_a=\rho_b=\rho_c$. The black point is a special point, where $\rho_a=\rho_b=\rho_c=(4-\lambda^2)/(2\lambda)$. The thick black line corresponds to $\rho_b=\rho_c=(4-\lambda^2)/(2\lambda)$.}
\end{figure}

\begin{figure}[H]
\centering
\begin{subfigure}{0.3\textwidth}
  \centering
  \includegraphics[width=\textwidth]{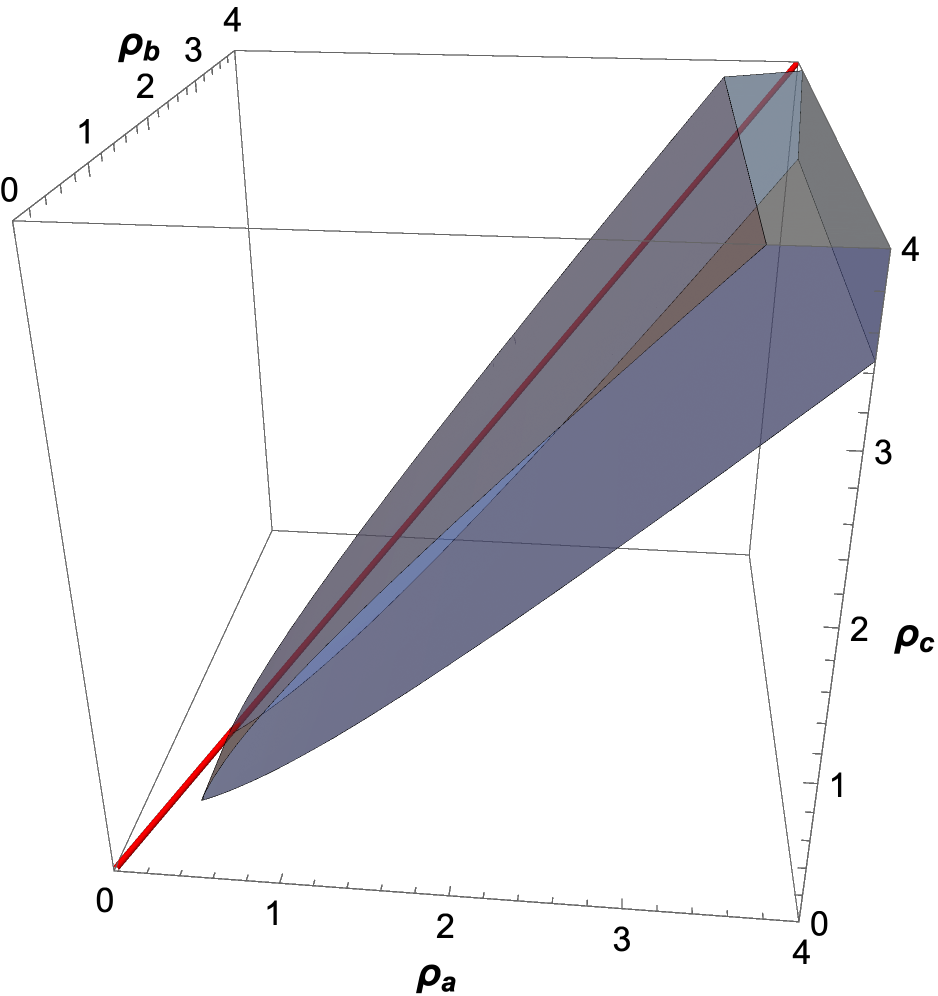}
  \caption{}
  \label{fig:Vec2xzStRegion}
\end{subfigure}
\qquad\qquad\qquad
\begin{subfigure}{0.3\textwidth}
  \centering
  \includegraphics[width=\textwidth]{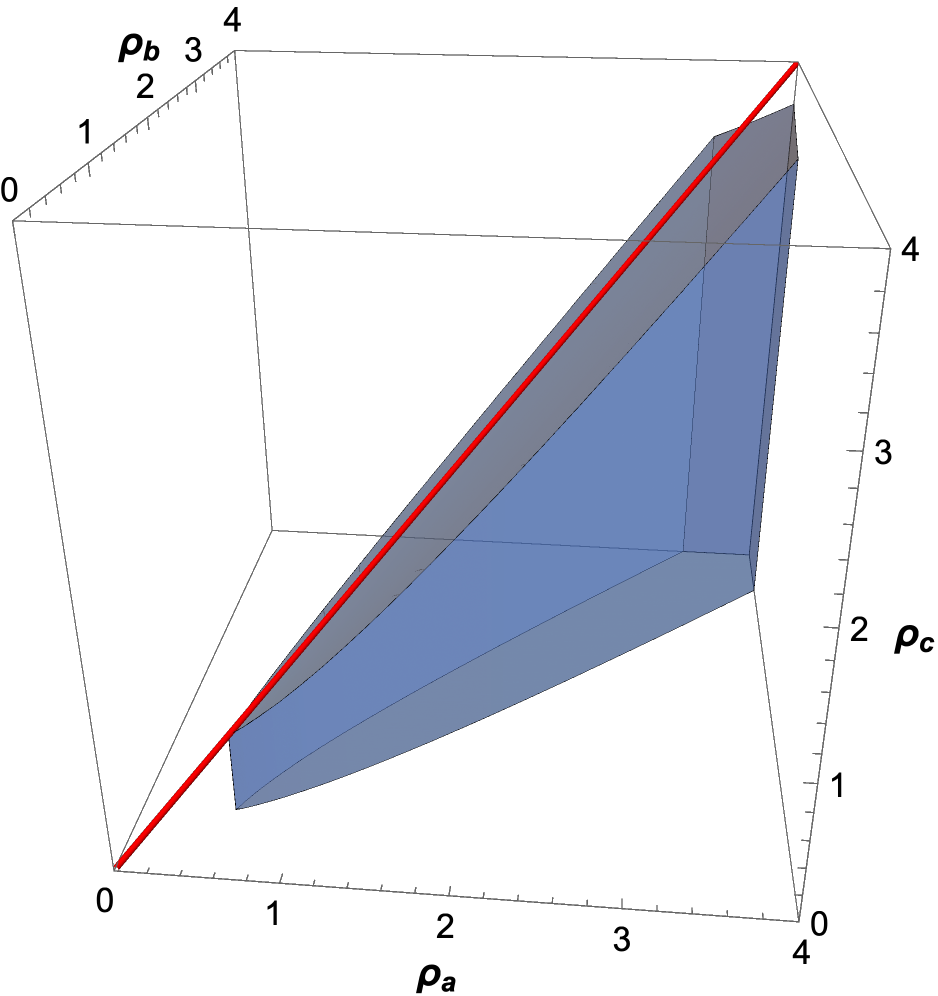}
  \caption{}
  \label{fig:Vec2xyStRegion}
\end{subfigure}
\caption{Stability regions of the fixed points type $\text{II}_{ac}$ (left) and $\text{II}_{ab}$ (right) colored as dark blue for $\lambda=1.5$.}
\end{figure}

Let us continue with the fixed point type $\text{II}_{bc}$. Its existence region, determined by Eq. \eqref{fixed point type IIbc ex cond}, is depicted in Fig. \ref{fig:Vec2yzExRegion} as the orange region, which is bounded by two surfaces: $-4+\lambda(\lambda+2\rho_b)-4\rho_c(\rho_c-\rho_b)=0$ and $-4+\lambda(\lambda+2\rho_c)-4\rho_b(\rho_b-\rho_c)=0$. It is noticeable that the existence region \eqref{fixed point type IIbc ex cond} will be determined by the following inequalities,
\begin{equation}
\begin{aligned}
-4+\lambda^2+2\lambda\rho_b-4\rho_c(\rho_c-\rho_b)&>0,\\
-4+\lambda^2+2\lambda\rho_c-4\rho_b(\rho_b-\rho_c)&>0,\\
\end{aligned}
\end{equation}
if $\lambda$, $\rho_a$, $\rho_b$, and $\rho_c$ are all positive. However, we have found that the stability region of this fixed point must follow the corresponding inequalities,
\begin{equation}\label{fixed point type IIbc st cond}
\begin{aligned}
-4+\lambda^2+2\lambda\rho_b-4\rho_c(\rho_c-\rho_b)&>0,\\
-4+\lambda^2+2\lambda\rho_c-4\rho_b(\rho_b-\rho_c)&>0,\\
-4+\lambda^2+2\lambda\rho_a-4\left(\Sigma_{\rho^2}-\rho_a\Sigma_\rho\right)&<0.
\end{aligned}
\end{equation}
It is apparent that the stability region differs from the existence region by the additional third inequality. To be more specific, this stability region is depicted in Fig. \ref{fig:Vec2yzStRegion} as the blue region, which is of course a part of the orange region shown in Fig. \ref{fig:Vec2yzExRegion}. Here, we notice that the surface $-4+\lambda^2+2\lambda\rho_a-4\left(\Sigma_{\rho^2}-\rho_a\Sigma_\rho\right)=0$ is also one of the surfaces that bound the stability region of the fixed point type III in Fig. \ref{fig:Vec3StRegion}. Interestingly, we see that the stability region of the fixed point type $\text{II}_{bc}$ is a thin layer surrounding the plane $\rho_b=\rho_c$, but the surface $-4+\lambda^2+2\lambda\rho_a-4\left(\Sigma_{\rho^2}-\rho_a\Sigma_\rho\right)=0$ prevents $\rho_a$ from getting too close to $\rho_b$ and $\rho_c$. Qualitatively speaking, therefore, $\rho_b$ and $\rho_c$ have to be sufficiently close to each other and significantly larger than $\rho_a$  in order to make the fixed point $\text{II}_{bc}$ stable. Moreover, the stability region terminates at the thick black line with $\rho_b=\rho_c=(4-\lambda^2)/(2\lambda)$ (note that it is identical to the thick black line in Fig. \ref{fig:Vec0StRegion}), which means that $\rho_b$ and $\rho_c$  have  to be sufficiently large. Similar arguments apply to the fixed points type $\text{II}_{ac}$ and $\text{II}_{ab}$. The stability regions of the fixed points type $\text{II}_{ac}$ and $\text{II}_{ab}$, depicted in Figs. \ref{fig:Vec2xzStRegion} and  \ref{fig:Vec2xyStRegion}, are obtained by exchanging $a\leftrightarrow b$ and $a\leftrightarrow c$ in \eqref{fixed point type IIbc st cond}, respectively.

%%%%%%%%%%%%%%%
\subsubsection{Fixed points type I}
\begin{figure}[H]
\centering
\begin{subfigure}{0.3\textwidth}
  \centering
  \includegraphics[width=\textwidth]{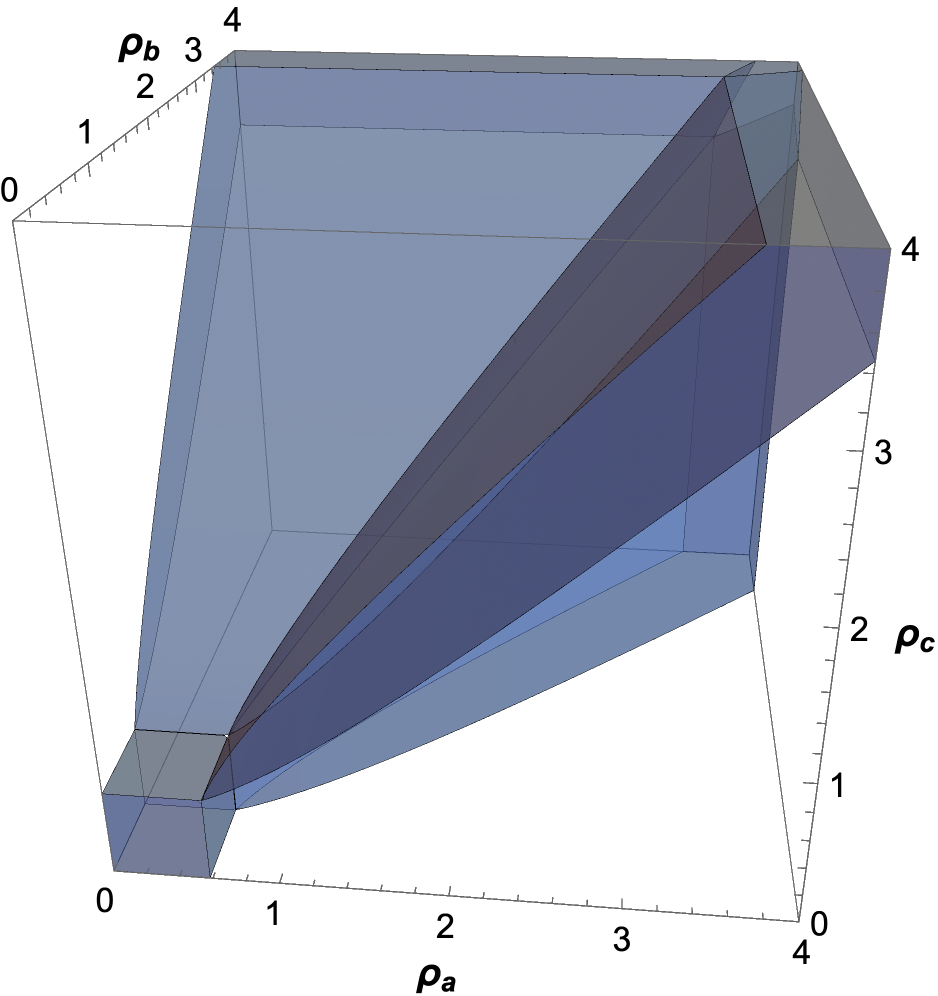}
  \caption{}
  \label{fig:Vec023StRegion a}
\end{subfigure}
\qquad\qquad\qquad
\begin{subfigure}{0.3\textwidth}
  \centering
  \includegraphics[width=\textwidth]{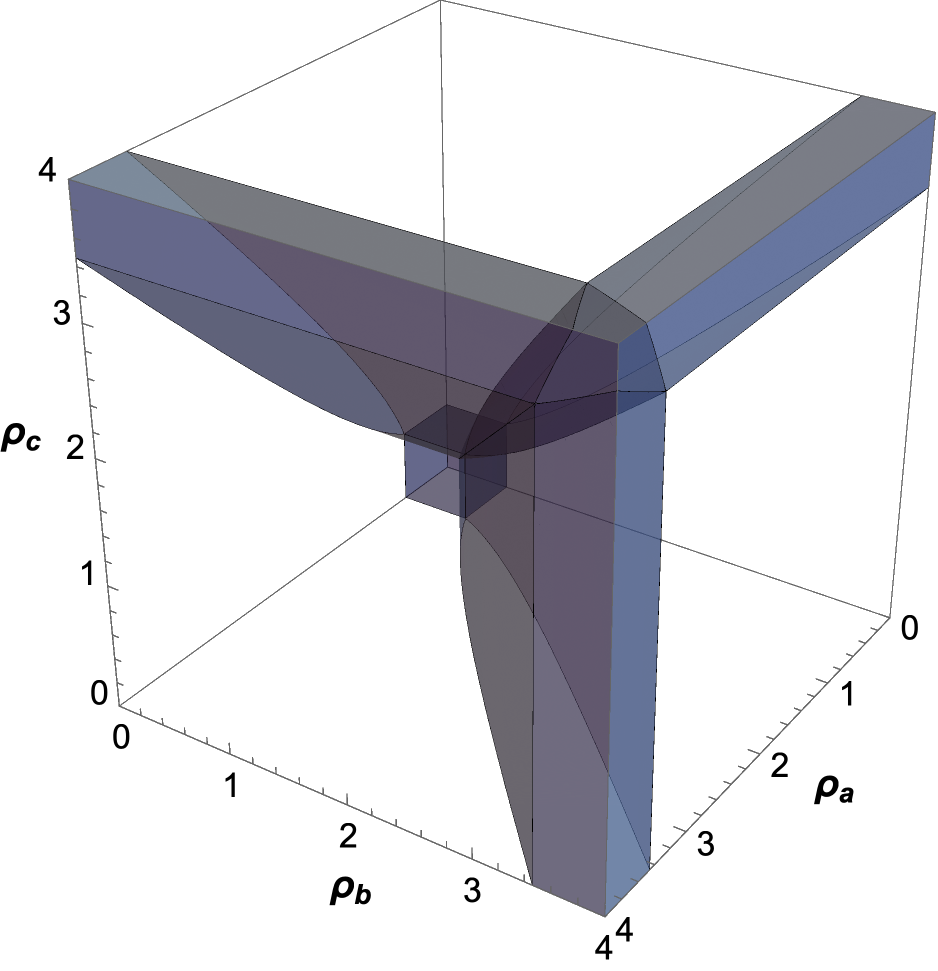}
  \caption{}
  \label{fig:Vec023StRegion b}
\end{subfigure}

\caption{Combination of the stability regions (colored as dark blue) of the fixed points type 0, III, $\text{II}_{bc}$, $\text{II}_{ac}$, and $\text{II}_{ab}$ for $\lambda=1.5$. The two figures are identical to each other but viewed from different angles.}
\label{fig:Vec023StRegion}
\end{figure}

\begin{figure}[H]
\centering

\begin{subfigure}{0.3\textwidth}
  \centering
  \includegraphics[width=\textwidth]{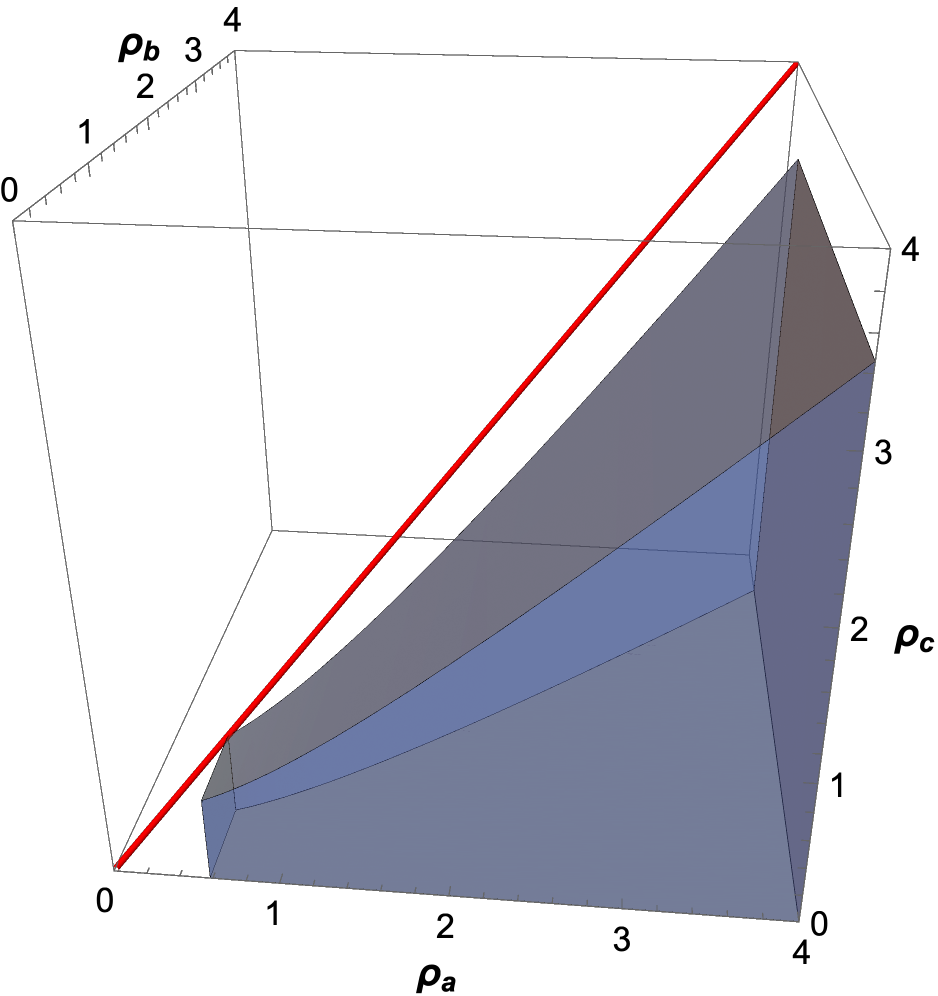}
  \caption{}
  \label{fig:Vec1xStRegion}
\end{subfigure}
\quad
\begin{subfigure}{0.3\textwidth}
  \centering
  \includegraphics[width=\textwidth]{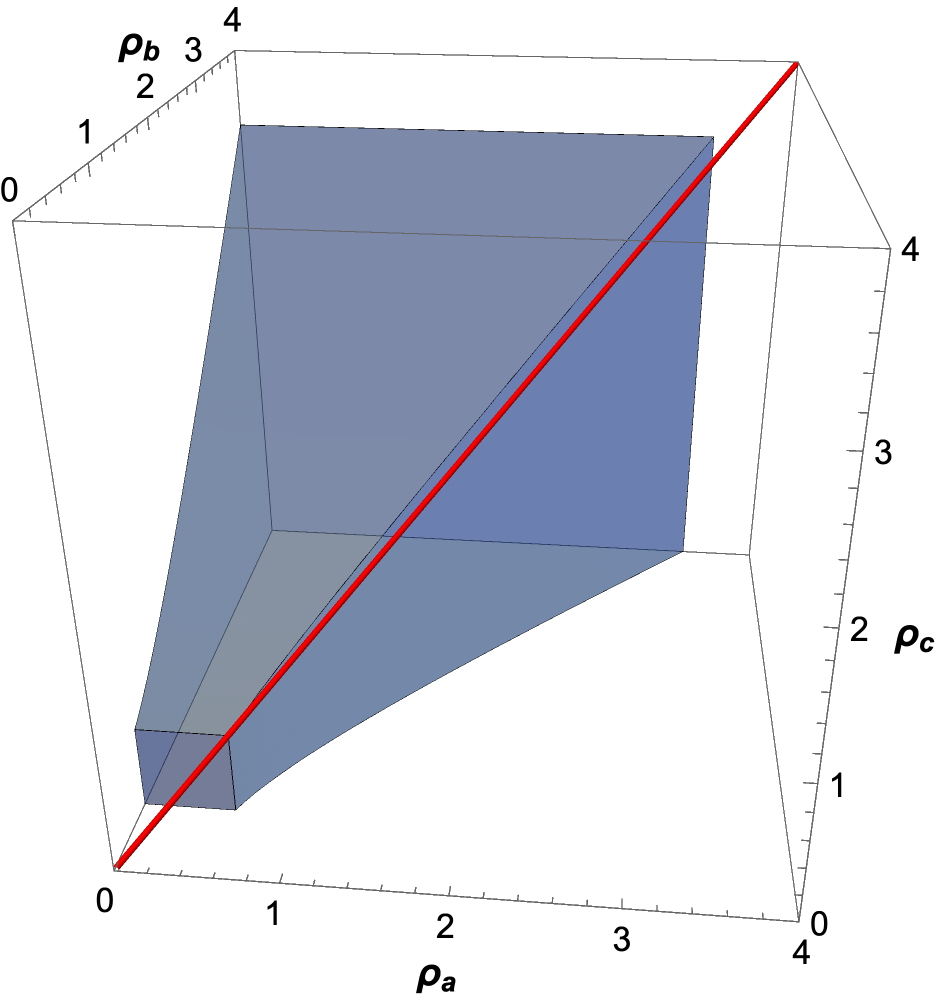}
  \caption{}
  \label{fig:Vec1yStRegion}
\end{subfigure}
\quad
\begin{subfigure}{0.3\textwidth}
  \centering
  \includegraphics[width=\textwidth]{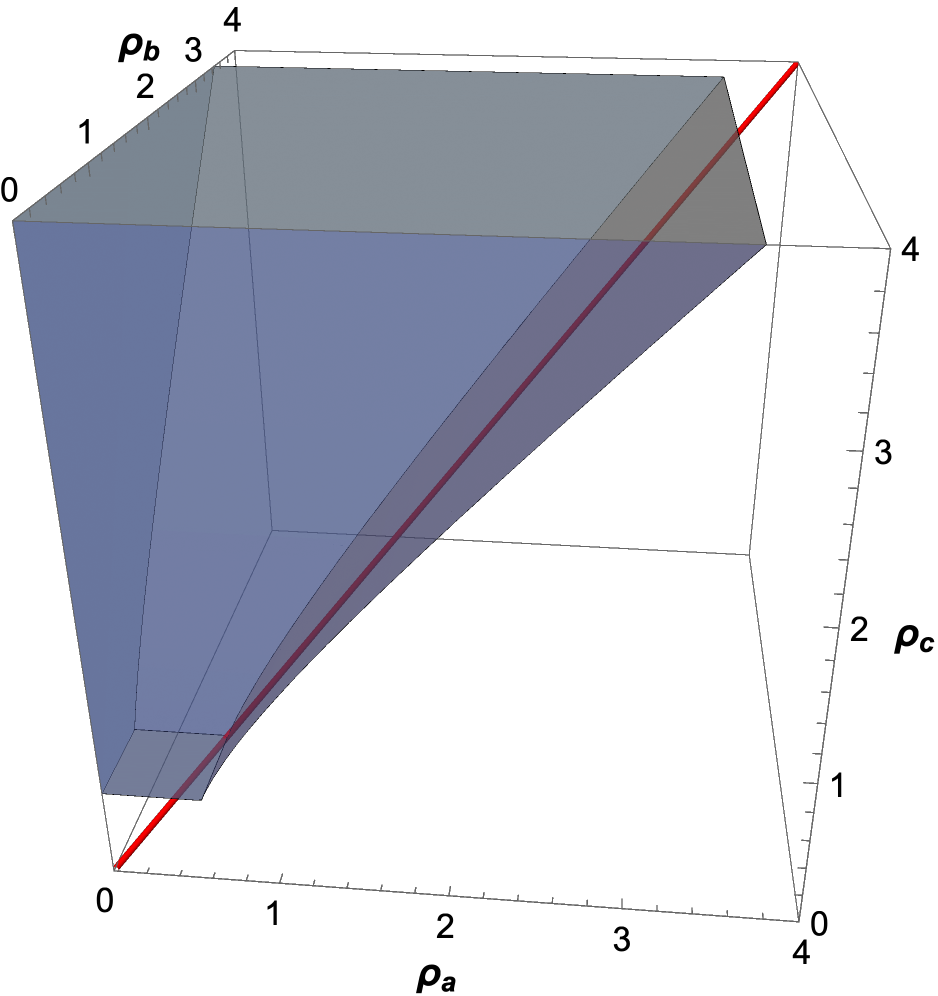}
  \caption{}
  \label{fig:Vec1zStRegion}
\end{subfigure}

\caption{Stability regions colored as dark blue of the fixed points type $\text{I}_a$ (left), $\text{I}_b$ (middle), and $\text{I}_c$ (right) for $\lambda=1.5$.}
\label{fig:Vec1StRegion}
\end{figure}

In the next step, we would like to combine the stability regions of the fixed points found above into one figure to form a \textit{three-bladed turbine-shaped} region (see Fig. \ref{fig:Vec023StRegion}). In this figure, the cube at the corner, the turbine shaft, and the turbine blades are the stability regions of the fixed points type 0, III, $\text{II}_{bc}$, $\text{II}_{ac}$, and $\text{II}_{ab}$, respectively. One might expect that the stability regions of the fixed points type $\text{I}_a$, $\text{I}_b$, and $\text{I}_c$ correspond to the spaces left between the blades. In fact, we found that this is the case. Let us focus on the stability region of the fixed point type $\text{I}_a$, which is determined by
\begin{equation}\label{fixed point type Ia st cond}
\begin{aligned}
-4+\lambda(\lambda+2\rho_b)-4\rho_a(\rho_a-\rho_b)<0,\\
-4+\lambda(\lambda+2\rho_c)-4\rho_a(\rho_a-\rho_c)<0,\\
\lambda^2+2\lambda\rho_a-4>0.
\end{aligned}
\end{equation}
and is depicted in Fig. \ref{fig:Vec1xStRegion} for $\lambda=1.5$. It is observed that this stability region is included in a region, in which $\rho_a>\rho_b$ and $\rho_a> \rho_c$. However, the surfaces $-4+\lambda(\lambda+2\rho_b)-4\rho_a(\rho_a-\rho_b)=0$ and $-4+\lambda(\lambda+2\rho_c)-4\rho_a(\rho_a-\rho_c)=0$ prevent $\rho_a$ from approaching too close to $\rho_b$ and $\rho_c$, respectively. Qualitatively speaking, therefore,  $\rho_a$ has to be significantly larger than $\rho_b$ and $\rho_c$ in order to make the fixed point type $\text{I}_a$ stable. Moreover, the stability region terminates at the plane $\rho_a=(4-\lambda^2)/(2\lambda)$, which means that $\rho_a$ has to be sufficiently large, too. Similar arguments apply to the fixed points type $\text{I}_b$ and $\text{I}_c$. The stability regions of the fixed points type $\text{I}_b$ and $\text{I}_c$, depicted in Figs. \ref{fig:Vec1yStRegion} and \ref{fig:Vec1zStRegion}, are obtained by exchanging $a\leftrightarrow b$ and $a\leftrightarrow c$ in \eqref{fixed point type Ia st cond}, respectively.

For convenience, we summarize the stability regions of the fixed points in Table \ref{tab:stability regions}.

\renewcommand{\arraystretch}{1.3}
\begin{table}[H]
\begin{ruledtabular}
\begin{tabular}{cccc}
\multirow[c]{2}{*}{\textbf{Fixed point type}} & \multicolumn{3}{c}{\textbf{Stability regions}} \\
\cline{2-4}
 & \textbf{Quantitative} & \textbf{Qualitative (for a fixed $\lambda$)} & \textbf{Illustration} \\
\hline
0 & \eqref{fixed point type 0 st cond} & \multicolumn{1}{l}{$\rho_a$, $\rho_b$, and $\rho_c$ are sufficiently small} & Fig. \ref{fig:Vec0StRegion} \\
\hline
\multirow[c]{2}{*}{$\text{I}_a$} & \multirow[c]{2}{*}{\eqref{fixed point type Ia st cond}} & \multicolumn{1}{l}{$\rho_a$ is sufficiently large} & \multirow[c]{2}{*}{Fig. \ref{fig:Vec1xStRegion}} \\
& & \multicolumn{1}{l}{$\rho_a$ is significantly larger than $\rho_b$ and $\rho_c$} \\
\hline
\multirow[c]{2}{*}{$\text{I}_b$} & \multirow[c]{2}{*}{\eqref{fixed point type Ia st cond} with $a\leftrightarrow b$} & \multicolumn{1}{l}{$\rho_b$ is sufficiently large} & \multirow[c]{2}{*}{Fig. \ref{fig:Vec1yStRegion}} \\
& & \multicolumn{1}{l}{$\rho_b$ is significantly larger than $\rho_a$ and $\rho_c$} \\
\hline
\multirow[c]{2}{*}{$\text{I}_c$} & \multirow[c]{2}{*}{\eqref{fixed point type Ia st cond} with $a\leftrightarrow c$} & \multicolumn{1}{l}{$\rho_c$ is sufficiently large} & \multirow[c]{2}{*}{Fig. \ref{fig:Vec1zStRegion}} \\
& & \multicolumn{1}{l}{$\rho_c$ is significantly larger than $\rho_a$ and $\rho_b$} \\
\hline
\multirow[c]{2}{*}{$\text{II}_{bc}$} & \multirow[c]{2}{*}{\eqref{fixed point type IIbc st cond}} & \multicolumn{1}{l}{$\rho_b$ and $\rho_c$ are sufficiently large and close} & \multirow[c]{2}{*}{Fig. \ref{fig:Vec2yzStRegion}} \\
& & \multicolumn{1}{l}{$\rho_b$ and $\rho_c$ are significantly larger than $\rho_a$} \\
\hline
\multirow[c]{2}{*}{$\text{II}_{ac}$} & \multirow[c]{2}{*}{\eqref{fixed point type IIbc st cond} with $a\leftrightarrow b$} & \multicolumn{1}{l}{$\rho_a$ and $\rho_c$ are sufficiently large and close} & \multirow[c]{2}{*}{Fig. \ref{fig:Vec2xzStRegion}} \\
& & \multicolumn{1}{l}{$\rho_a$ and $\rho_c$ are significantly larger than $\rho_b$} \\
\hline
\multirow[c]{2}{*}{$\text{II}_{ab}$} & \multirow[c]{2}{*}{\eqref{fixed point type IIbc st cond} with $a\leftrightarrow c$} & \multicolumn{1}{l}{$\rho_a$ and $\rho_b$ are sufficiently large and close} & \multirow[c]{2}{*}{Fig. \ref{fig:Vec2xyStRegion}} \\
& & \multicolumn{1}{l}{$\rho_a$ and $\rho_b$ are significantly larger than $\rho_c$} \\
\hline
III & \eqref{fixed point type III ex cond} & \multicolumn{1}{l}{$\rho_a$, $\rho_b$, and $\rho_c$ are sufficiently large and close} & Fig. \ref{fig:Vec3StRegion} \\
\end{tabular}
\end{ruledtabular}
\caption{\label{tab:stability regions} Summary of the stability regions of the fixed points.}
\end{table}

Furthermore, we can combine the results presented in  Tables \ref{tab: power-law solutions} and \ref{tab:stability regions} to draw a general qualitative conclusion on the fate of the vector fields and the metric of spacetime as follows. If all the coupling constants are sufficiently small, the vector fields are eventually diluted, regardless of their relative values with respect to each other. But if at least one of the coupling constants is sufficiently large, the vector field with the largest coupling constant, denoted by $\rho_\text{max}$, persists as the universe expands. Moreover, any vector field with a coupling constant that is smaller but sufficiently close to $\rho_\text{max}$ persists as well. However, any vector field with a coupling constant that is significantly smaller than $\rho_\text{max}$ is eventually diluted. On the other hand, the fate of spacetime's metric depends on the number of persisting vector fields. If there are three or two persisting vector fields, the metric of spacetime in general approaches a general power-law BI metric. If only one vector field persists, the metric approaches a power-law rsBI metric. If all the vector fields are diluted, then the metric approaches a power-law FLRW metric, which is consistent with the cosmic no-hair conjecture.
%%%%%%%%%%%%%%%%%%%%
\subsection{Numerical analysis}
\begin{figure}[H]
\centering
\begin{subfigure}{0.3\textwidth}
  \centering
  \includegraphics[width=\textwidth]{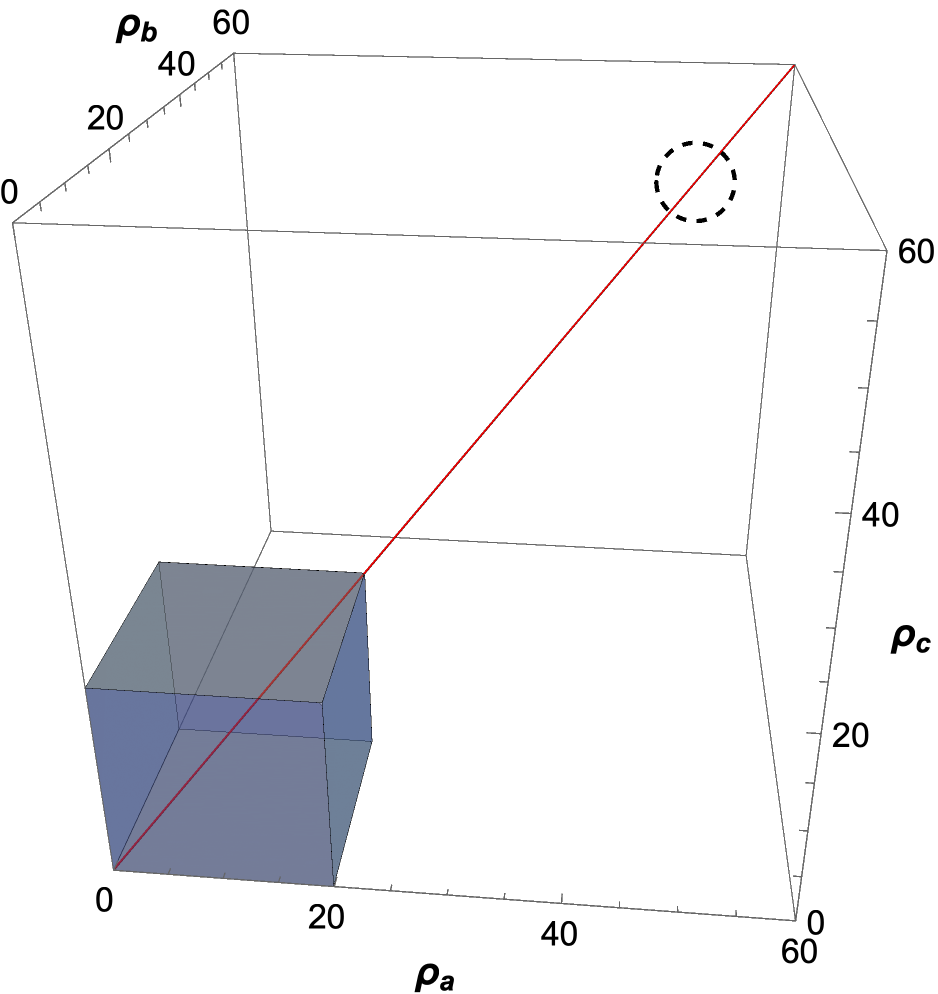}
  \caption{}
  \label{fig:Vec0StRegionInf}
\end{subfigure}
\qquad\qquad\qquad
\begin{subfigure}{0.3\textwidth}
  \centering
  \includegraphics[width=\textwidth]{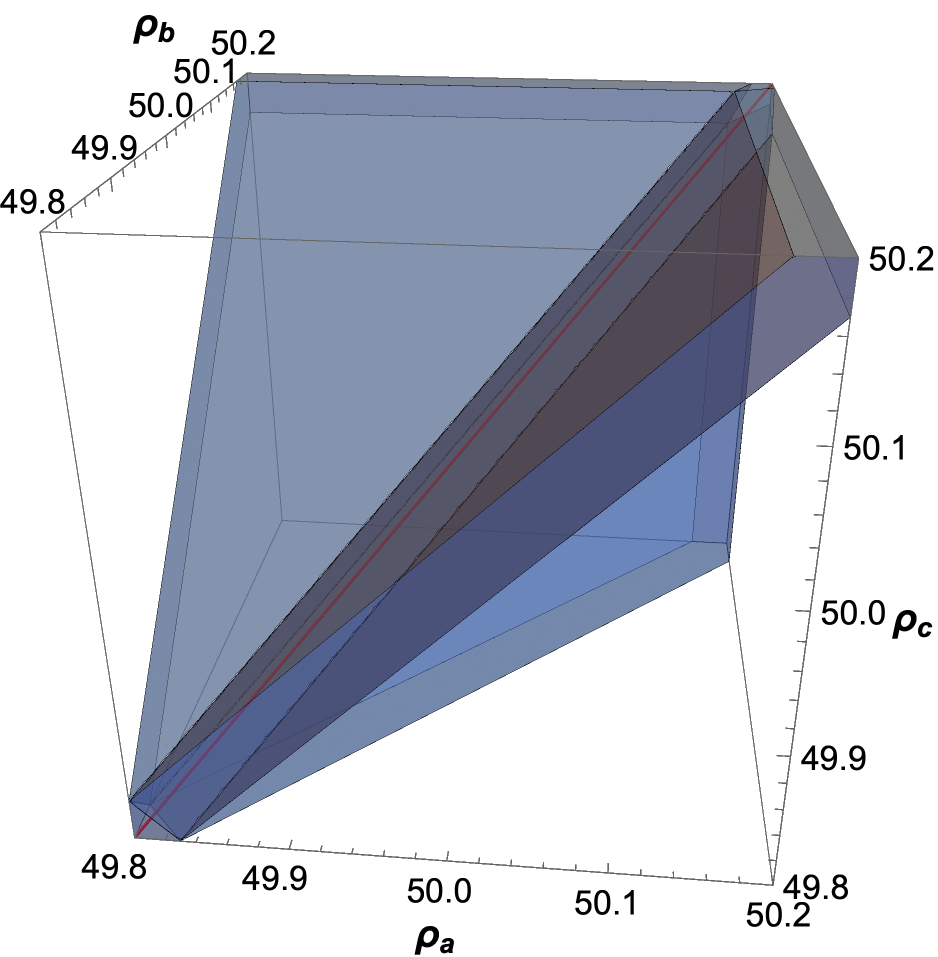}
  \caption{}
  \label{fig:Vec023StRegionInf}
\end{subfigure}

\caption{(left) Stability region of the fixed point type 0 and (right) a combination of stability regions of the fixed point types III, $\text{II}_{bc}$, $\text{II}_{ac}$, and $\text{II}_{ab}$ for $\lambda=0.1$. The right figure is the enlarged version of the dashed circle region in the left figure.}
\label{fig:StRegionInf}
\end{figure}

To provide strong support for our above arguments on the stability of fixed points, we would like to solve the dynamical system of autonomous equations \eqref{autonomous eq Xy}-\eqref{autonomous eq Zz} numerically \cite{Numerical analysis}. However, we should first note that the above stability analysis is general enough to be applied to arbitrary expansion rate and spatial anisotropies. The above choice $\lambda=1.5$ above is merely for illustration. To realize a realistic anisotropic inflation, we need $\zeta$ to be very large and the spatial anisotropies to be very small compared to $\zeta$ \cite{Kanno:2010nr}. We therefore focus on the value $\lambda=0.1$ for numerical calculations, following the previous study in Ref. \cite{Kanno:2010nr}, while the values of the coupling constants $\rho_i$ will be appropriately chosen for each considered fixed point. The stability region of the fixed point type 0 is depicted in Fig. \ref{fig:Vec0StRegionInf}. Unfortunately, the stability regions of the other fixed points are very difficult to distinguish when displayed in Fig. \ref{fig:Vec0StRegionInf}. Therefore, we depict the stability regions of the fixed points type $\text{II}_{bc}$, $\text{II}_{ac}$, $\text{II}_{ab}$, and III in Fig.  \ref{fig:Vec023StRegionInf}, where the coupling constants are restricted between $49.8$ and $50.2$. It is noted that Figs.  \ref{fig:Vec0StRegionInf} and \ref{fig:Vec023StRegionInf} are analogous to Fig.  \ref{fig:Vec023StRegion a}. We will not show the stability regions of the fixed points $\text{I}_{a}$, $\text{I}_{b}$, and $\text{I}_{c}$ for $\lambda=0.1$ since they are analogous to Figs. \ref{fig:Vec1xStRegion}, \ref{fig:Vec1yStRegion}, and \ref{fig:Vec1zStRegion}, respectively.

\begin{figure}[H]
\centering

\begin{subfigure}{0.31\textwidth}
  \centering
  \includegraphics[width=\textwidth]{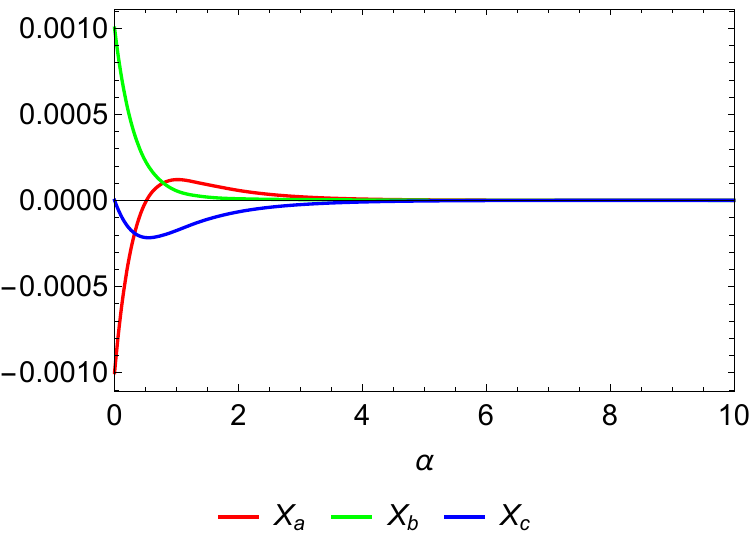}
  \caption{}
  \label{fig:Vec0ConvInfX}
\end{subfigure}
\quad
\begin{subfigure}{0.31\textwidth}
  \centering
  \includegraphics[width=\textwidth]{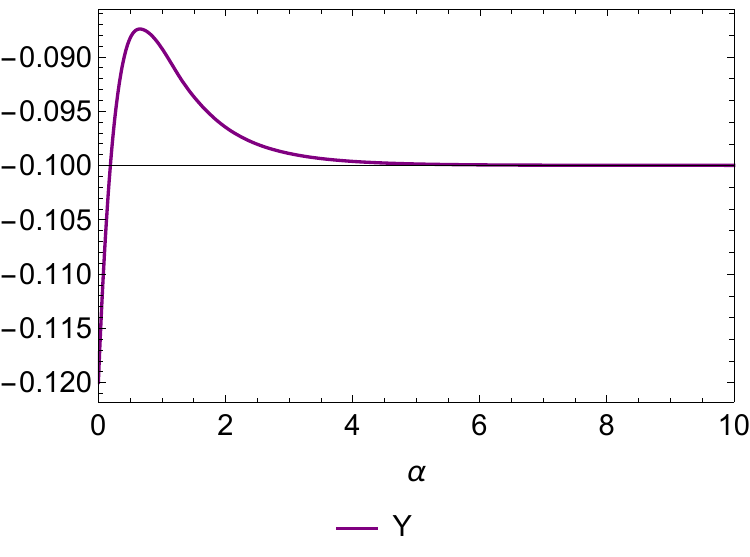}
  \caption{}
  \label{fig:Vec0ConvInfY}
\end{subfigure}
\quad
\begin{subfigure}{0.31\textwidth}
  \centering
  \includegraphics[width=\textwidth]{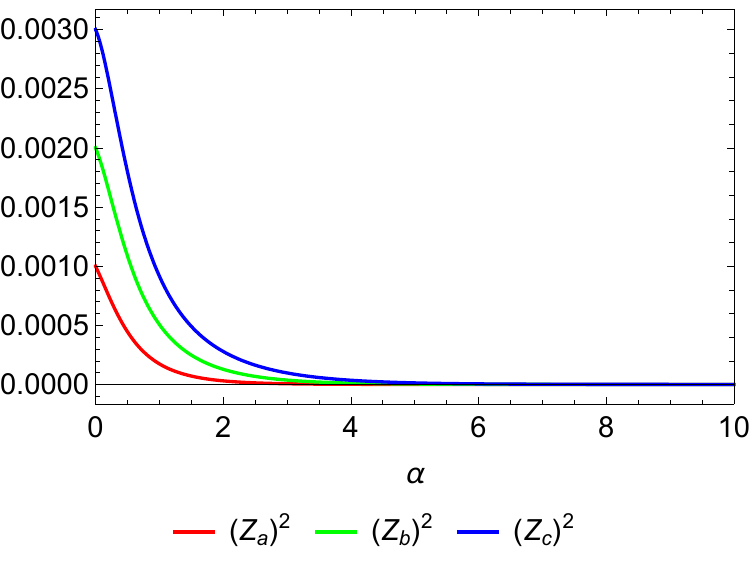}
  \caption{}
  \label{fig:Vec0ConvInfZ2}
\end{subfigure}

\caption{Convergence of the dynamical variables to the fixed point type 0 for $\lambda=0.1$, $\rho_a=12$, $\rho_b=14$, and $\rho_c=15$.}
\label{fig:Vec0ConvInf}
\end{figure}

To verify the attractor behavior of the fixed points, we numerically solved the autonomous system for several sets of initial conditions. However, in the following figures, we only present one representative set of initial conditions for each fixed point. We start the numerical analysis with the fixed point type 0 by choosing $\lambda=0.1$, $\rho_a=12$, $\rho_b=14$, and $\rho_c=15$, which satisfy the stability conditions shown in Eq. \eqref{fixed point type 0 st cond} ($\rho_a$, $\rho_b$, and $\rho_c$ are sufficiently small). The fixed point is therefore defined to be
\begin{equation}\label{Vec0 numerical fixed point}
\begin{aligned}
X_a=X_b=X_c=0,\quad Y=-0.1,\quad Z_a=Z_b=Z_c=0.
\end{aligned}
\end{equation}
At the fixed point \eqref{Vec0 numerical fixed point}, the anisotropies vanish and $\zeta=200\gg 1$ according to the solution \eqref{power-law solution type 0}. In Fig. \ref{fig:Vec0ConvInf}, we see that the dynamical variables $X_a$, $X_b$, $X_c$, $Y$, $(Z_a)^2$, $(Z_b)^2$, and $(Z_c)^2$ all tend to converge to the fixed point \eqref{Vec0 numerical fixed point} as expected. This confirms that the fixed point type 0 is an attractor. 

\begin{figure}[H]
\centering

\begin{subfigure}{0.31\textwidth}
  \centering
  \includegraphics[width=\textwidth]{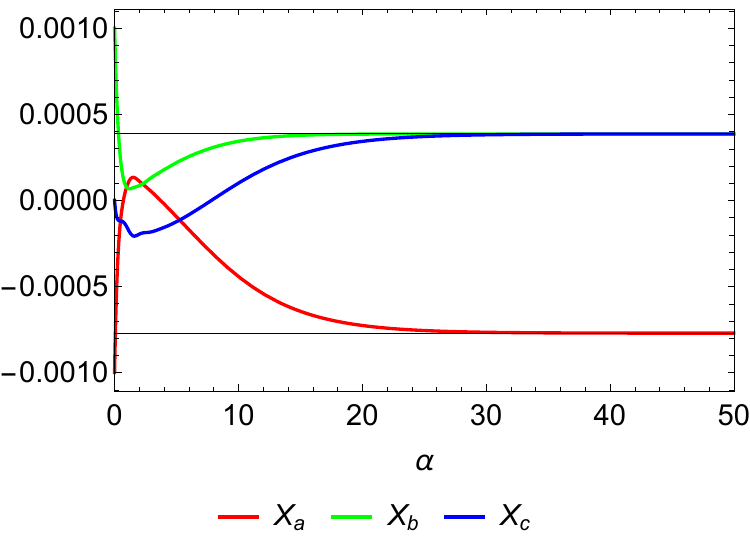}
  \caption{}
  \label{fig:Vec1xConvInfX}
\end{subfigure}
\quad
\begin{subfigure}{0.31\textwidth}
  \centering
  \includegraphics[width=\textwidth]{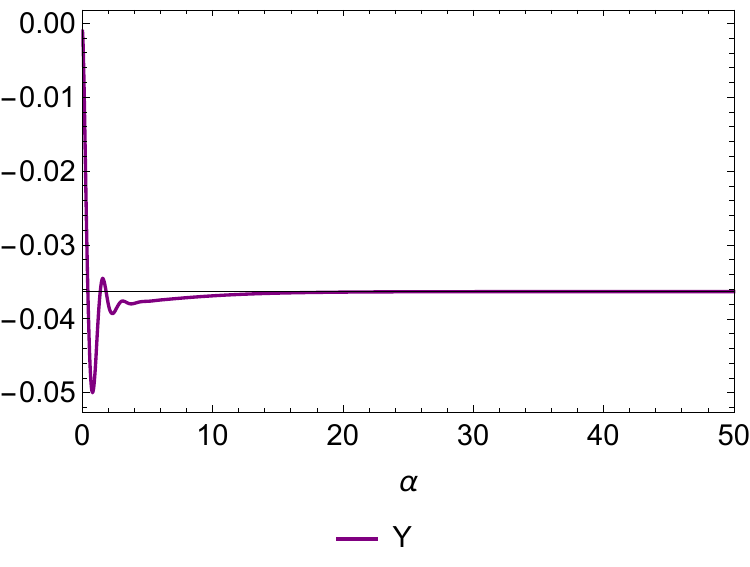}
  \caption{}
  \label{fig:Vec1xConvInfY}
\end{subfigure}
\quad
\begin{subfigure}{0.31\textwidth}
  \centering
  \includegraphics[width=\textwidth]{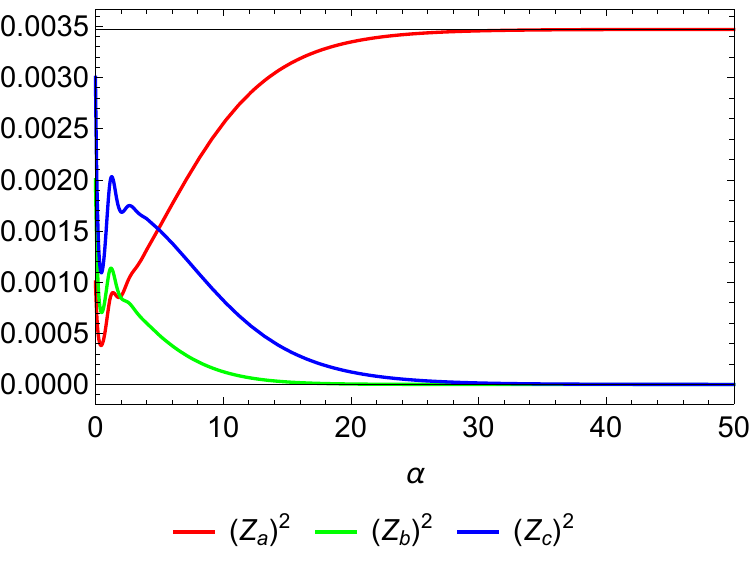}
  \caption{}
  \label{fig:Vec1xConvInfZ2}
\end{subfigure}

\caption{Convergence of the dynamical variables to the fixed point type $\text{I}_a$ for $\lambda=0.1$, $\rho_a=55$, $\rho_b=50$, and $\rho_c=52$.}
\label{fig:Vec1xConvInf}
\end{figure}

Next, we move on to the fixed point type $\text{I}_a$ by choosing $\lambda=0.1$, $\rho_a=55$, $\rho_b=50$, and $\rho_c=52$, which satisfy the corresponding stability conditions shown in Eq. \eqref{fixed point type Ia st cond} ($\rho_a$ is sufficiently large and significantly larger than $\rho_b$ and $\rho_c$). The fixed point is therefore given by
\begin{equation}\label{Vec1x numerical fixed point}
\begin{aligned}
X_a&\approx-0.0007713,\quad X_b=X_c\approx 0.0003857,\quad \\
Y&\approx-0.03634,\quad (Z_a)^2\approx 0.003469,\quad (Z_b)^2=(Z_c)^2=0.
\end{aligned}
\end{equation}
For the fixed point \eqref{Vec1x numerical fixed point}, the spatial anisotropies are very small and $\zeta\approx 550\gg 1$ according to the solution \eqref{power-law solution type Ia}. In Fig. \ref{fig:Vec1xConvInf}, we see that the dynamical variables $X_a$, $X_b$, $X_c$, $Y$, $(Z_a)^2$, $(Z_b)^2$, and $(Z_c)^2$ converge to the fixed point \eqref{Vec1x numerical fixed point} as expected. Hence, we can conclude that this fixed point is an attractor. Similar results can be straightforwardly obtained for the fixed points type $\text{I}_b$ and $\text{I}_c$. 

\begin{figure}[H]
\centering

\begin{subfigure}{0.31\textwidth}
  \centering
  \includegraphics[width=\textwidth]{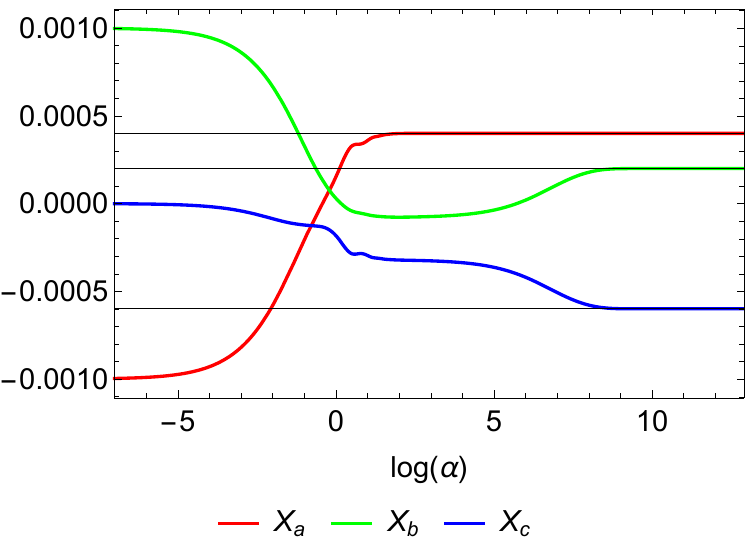}
  \caption{}
  \label{fig:Vec2yzConvInfX}
\end{subfigure}
\quad
\begin{subfigure}{0.31\textwidth}
  \centering
  \includegraphics[width=\textwidth]{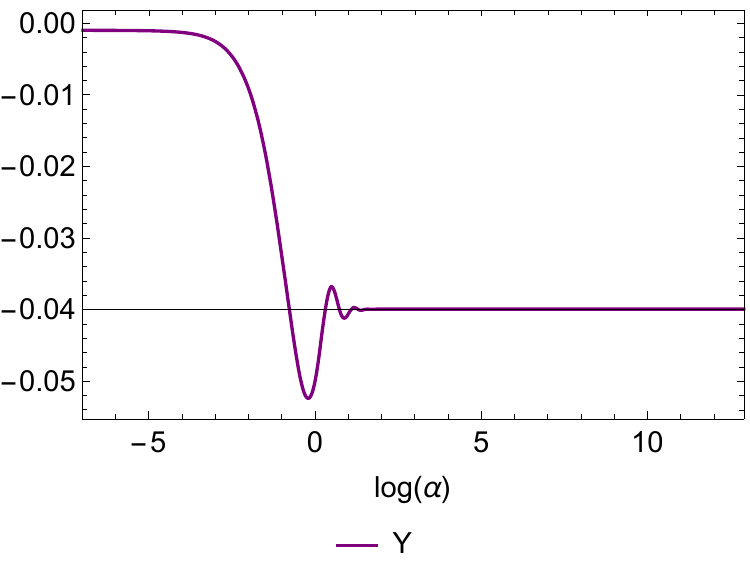}
  \caption{}
  \label{fig:Vec2yzConvInfY}
\end{subfigure}
\quad
\begin{subfigure}{0.31\textwidth}
  \centering
  \includegraphics[width=\textwidth]{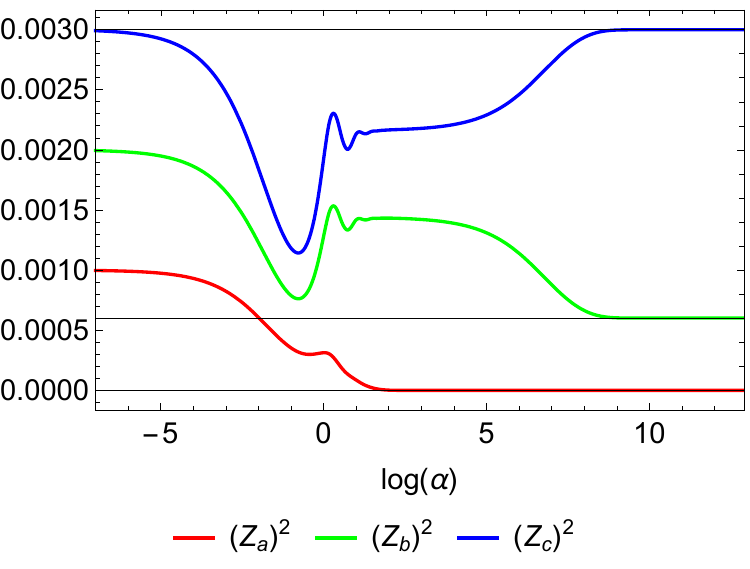}
  \caption{}
  \label{fig:Vec2yzConvInfZ2}
\end{subfigure}

\caption{Convergence of the dynamical variables to the fixed point type $\text{II}_{bc}$ for $\lambda=0.1$, $\rho_a=40$, $\rho_b=50$, and $\rho_c=50.02$.}
\label{fig:Vec2yzConvInf}
\end{figure}

We continue with the fixed point type $\text{II}_{bc}$. We choose $\lambda=0.1$, $\rho_a=40$, $\rho_b=50$, and $\rho_c=50.02$, which satisfy the corresponding stability conditions shown in Eq. \eqref{fixed point type IIbc st cond} ($\rho_b$ and $\rho_c$ are sufficiently large, close to each other, and significantly larger than $\rho_a$). The fixed point therefore takes the following value,
\begin{equation}\label{Vec2yz numerical fixed point}
\begin{aligned}
X_a&\approx 0.0004000,\quad X_b\approx 0.0001995,\quad X_c\approx-0.0005996,\\
Y&\approx-0.03996,\quad (Z_a)^2=0,\quad (Z_b)^2\approx 0.000601,\quad (Z_c)^2\approx 0.002997.
\end{aligned}
\end{equation}
For the fixed point \eqref{Vec2yz numerical fixed point}, the spatial anisotropies are very small and $\zeta\approx 500 \gg 1$ according to the solution \eqref{power-law solution type IIbc}. In Fig. \ref{fig:Vec2yzConvInf}, we see that the dynamical variables $X_a$, $X_b$, $X_c$, $Y$, $(Z_a)^2$, $(Z_b)^2$, and $(Z_c)^2$ tend to converge to the fixed point \eqref{Vec2yz numerical fixed point} as expected, confirming the attractor property of the fixed point type $\text{II}_{bc}$. Note that we have chosen $\log(\alpha)$ instead of $\alpha$ for the horizontal axis for better visual clarity. Similar results can be obtained straightforwardly for the fixed points type $\text{II}_{ac}$ and $\text{II}_{ab}$. 

\begin{figure}[H]
\centering

\begin{subfigure}{0.31\textwidth}
  \centering
  \includegraphics[width=\textwidth]{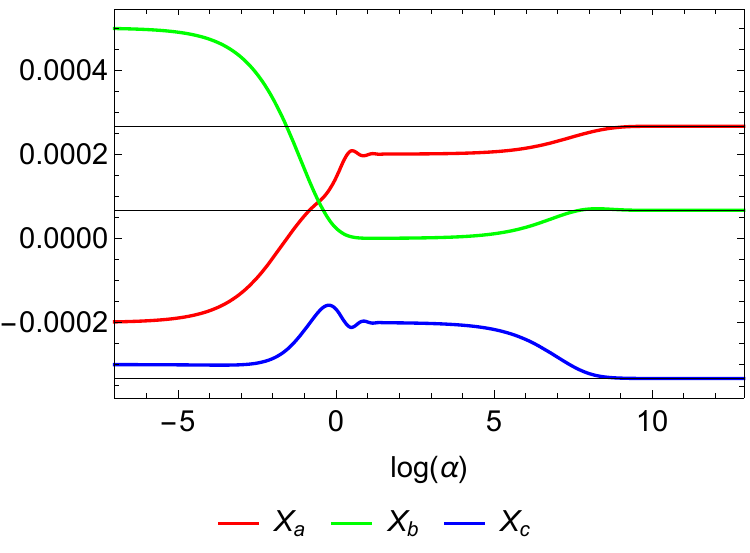}
  \caption{}
  \label{fig:Vec3ConvInfX}
\end{subfigure}
\quad
\begin{subfigure}{0.31\textwidth}
  \centering
  \includegraphics[width=\textwidth]{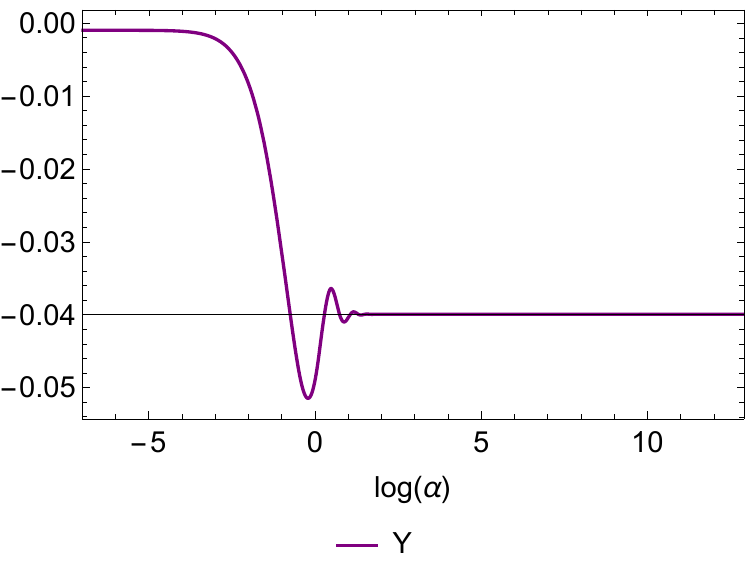}
  \caption{}
  \label{fig:Vec3ConvInfY}
\end{subfigure}
\quad
\begin{subfigure}{0.31\textwidth}
  \centering
  \includegraphics[width=\textwidth]{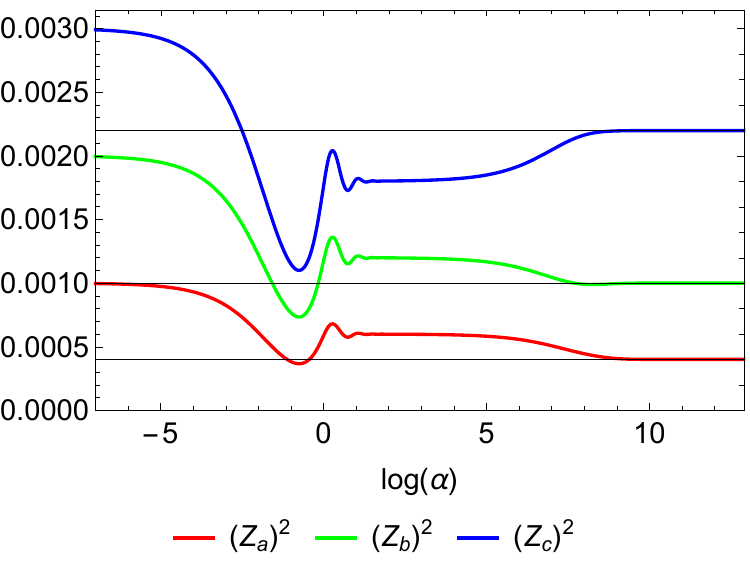}
  \caption{}
  \label{fig:Vec3ConvInfZ2}
\end{subfigure}

\caption{Convergence of the dynamical variables to the fixed point type III for $\lambda=0.1$, $\rho_a=49.995$, $\rho_b=50$, and $\rho_c=50.01$.}
\label{fig:Vec3ConvInf}
\end{figure}

For the fixed point type III, we choose $\lambda=0.1$, $\rho_a=49.995$, $\rho_b=50$, and $\rho_c=50.01$, which clearly satisfy the stability conditions shown in Eq.  \eqref{fixed point type III ex cond} ($\rho_a$, $\rho_b$, and $\rho_c$ are sufficiently large and close to each other). The fixed point therefore reads
\begin{equation}\label{Vec3 numerical fixed point}
\begin{aligned}
X_a&\approx 0.0002664,\quad X_b\approx 0.0000666,\quad X_c\approx-0.0003330,\quad Y\approx-0.03996,\\
(Z_a)^2&\approx 0.000401,\quad (Z_b)^2\approx 0.001000,\quad (Z_c)^2\approx 0.002198.
\end{aligned}
\end{equation}
Also, for the fixed point \eqref{Vec3 numerical fixed point}, the spatial anisotropies are very small and $\zeta\approx 500 \gg 1$ according to the solution \eqref{power-law solution vec3}. In Fig. \ref{fig:Vec3ConvInf}, we see that the dynamical variables $X_a$, $X_b$, $X_c$, $Y$, $(Z_a)^2$, $(Z_b)^2$, and $(Z_c)^2$ all tend to converge to the fixed point \eqref{Vec3 numerical fixed point} as expected, displaying the attractor property of the fixed point type III.

%%%%%%%%%%%%%%%%%
\section{Conclusions}\label{sec5}
We have investigated a model of Bianchi type I inflationary universe where the inflaton field is non-minimally coupled to three vector fields aligned along three axes. As a result, we found four types of power-law solutions (types 0, I, II, and III) that are classified according to the number of non-vanishing vector fields.  All these solutions have been summarized in the Table \ref{tab: power-law solutions}. By investigating stability analysis, we have shown that all obtained solutions can be attractors and stable against perturbations under specific conditions for the stability regions. It has turned out that the stability properties of the obtained solutions strongly depend on the values of the coupling constants relative to the others. This dependence has been described in Sect. \ref{sec4} both quantitatively and qualitatively.  To have a systematic view of the stability of the derived solutions, we have built Table \ref{tab:stability regions} listing all possible cases. The attractor properties of these solutions have been confirmed by numerical calculations. 

One might ask whether our analysis can be extended to other classes of anisotropic cosmologies, such as different Bianchi models. Interestingly, stability of Bianchi models with a single field has been investigated in Ref. \cite{Hervik:2011xm,Normann:2017aav}. For multiple fields, a crucial step is to construct field configurations consistent with the underlying metrics. In the Bianchi type I case, we considered three orthogonal vector fields aligned with the spatial axes. By contrast, for other Bianchi types, identifying suitable configurations of multiple vector fields may be less straightforward. Nevertheless, once such configurations are established, the same procedure presented in this work can be applied to investigate the evolution and fate of both vector fields and anisotropies. We hope that this direction of research will be further explored in future works.

Since our present study is merely focused on the classification of all possible stable solutions of a generalized KSW model, in which we maximize the number of vector fields as well as the number of spatial anisotropies, we have not discussed their CMB imprints. This issue will be left for future studies. For now,  we hope that our results will be useful for other studies on the early universe in general and the evolution of anisotropies and vector fields during inflation in particular.  For example, a recent paper \cite{Kouniatalis:2025qfz} has proposed a non-trivial connection between the wave function of the universe and cosmic inflation. It would therefore be interesting to test the validity of the cosmic no-hair conjecture in this scenario. Another possible direction for future work is to extend our analysis to a novel inflaton model with a generalized exponential plateau proposed in Ref. \cite{Kouniatalis:2025orn}.
%%%%%%%%%%%%%%%%%%%%%%
\begin{acknowledgments}
This study is funded by the Vietnam National Foundation for Science and Technology Development (NAFOSTED) under grant number 103.01-2023.50. We are grateful to the anonymous referees for their constructive comments and suggestions. We thank Prof. P. V. Dong very much for his valuable support. We also thank G.~Kouniatalis very much for his useful discussions.
\end{acknowledgments}
%%%%%%%%%%%%%%%
\appendix
\section{Autonomous equations and fixed points} \label{Appendix 1}
Armed with the dynamical variables defined in Eq. \eqref{dynamical variables}, we can rewrite the field equations \eqref{Friedmann eq 1}, \eqref{Friedmann eq 2}, \eqref{Friedmann eq 3}, \eqref{Friedmann eq 4}, and \eqref{scalar field eq} as follows
\begin{align}
\frac{V}{\dot{\alpha}^2}&=3-(X_b)^2-(X_c)^2-X_b X_c-\frac{Y^2}{2}-\frac{(Z_a)^2}{2}-\frac{(Z_b)^2}{2}-\frac{(Z_c)^2}{2}>0,\label{Hamiltonian constraint}\\
\frac{\ddot{\alpha}}{\dot{\alpha}^2}&=-(X_b)^2-(X_c)^2-X_b X_c-\frac{Y^2}{2}-\frac{(Z_a)^2}{3}-\frac{(Z_b)^2}{3}-\frac{(Z_c)^2}{3},\label{Friedmann eq 2 in terms of dynamical variables}\\
\frac{\ddot{\sigma}_b}{\dot{\alpha}^2}&=-3X_b+\frac{(Z_a)^2}{3}-\frac{2(Z_b)^2}{3}+\frac{(Z_c)^2}{3},\label{Friedmann eq 3 in terms of dynamical variables}\\
\frac{\ddot{\sigma}_c}{\dot{\alpha}^2}&=-3X_c+\frac{(Z_a)^2}{3}+\frac{(Z_b)^2}{3}-\frac{2(Z_c)^2}{3},\label{Friedmann eq 4 in terms of dynamical variables}\\
\frac{\ddot{\phi}}{\dot{\alpha}^2}&=\lambda\left[-3+(X_b)^2+(X_c)^2+X_b X_c+\frac{Y^2}{2}\right]-3Y+\left(\frac{\lambda}{2}+\rho_a\right)(Z_a)^2+\left(\frac{\lambda}{2}+\rho_b\right)(Z_b)^2+\left(\frac{\lambda}{2}+\rho_c\right)(Z_c)^2.\label{scalar field eq in terms of dynamical variables}
\end{align}
where the inequality has been added in \eqref{Hamiltonian constraint} to ensure the positivity of the potential. On the other hand, the autonomous equations are defined as follows
\begin{align}
\frac{dX_b}{d\alpha}&=\frac{1}{\dot{\alpha}} \frac{d}{dt}\left(\frac{\dot{\sigma}_b}{\dot{\alpha}}\right)=\frac{\ddot{\sigma}_b}{\dot{\alpha}^2}-X_b\frac{\ddot{\alpha}}{\dot{\alpha}^2},\label{dXb/d alpha}\\
\frac{dX_c}{d\alpha}&=\frac{1}{\dot{\alpha}} \frac{d}{dt}\left(\frac{\dot{\sigma}_c}{\dot{\alpha}}\right)=\frac{\ddot{\sigma}_c}{\dot{\alpha}^2}-X_c\frac{\ddot{\alpha}}{\dot{\alpha}^2},\label{dXc/d alpha}\\
\frac{dY}{d\alpha}&=\frac{1}{\dot{\alpha}} \frac{d}{dt}\left(\frac{\dot{\phi}}{\dot{\alpha}}\right)=\frac{\ddot{\phi}}{\dot{\alpha}^2}-Y\frac{\ddot{\alpha}}{\dot{\alpha}^2},\label{dY/d alpha}\\
\frac{dZ_a}{d\alpha}&=\frac{1}{\dot{\alpha}} \frac{d}{dt}\left(\frac{p_a f_a^{-1}}{\dot{\alpha}}e^{-2\alpha-\sigma_b-\sigma_c}\right)=-Z_a\left(2+X_b+X_c+\rho_a Y+\frac{\ddot{\alpha}}{\dot{\alpha}^2}\right),\label{dZa/d alpha}\\
\frac{dZ_b}{d\alpha}&=\frac{1}{\dot{\alpha}} \frac{d}{dt}\left(\frac{p_b f_b^{-1}}{\dot{\alpha}}e^{-2\alpha+\sigma_b}\right)=-Z_b\left(2-X_b+\rho_{b} Y+\frac{\ddot{\alpha}}{\dot{\alpha}^2}\right),\label{dZb/d alpha}\\
\frac{dZ_c}{d\alpha}&=\frac{1}{\dot{\alpha}} \frac{d}{dt}\left(\frac{p_c f_c^{-1}}{\dot{\alpha}}e^{-2\alpha+\sigma_c}\right)=-Z_c\left(2-X_c+\rho_{c} Y+\frac{\ddot{\alpha}}{\dot{\alpha}^2}\right). \label{dZc/d alpha}
\end{align}
Substituting Eqs. \eqref{Friedmann eq 2 in terms of dynamical variables}, \eqref{Friedmann eq 3 in terms of dynamical variables}, \eqref{Friedmann eq 4 in terms of dynamical variables}, and \eqref{scalar field eq in terms of dynamical variables} into Eqs. \eqref{dXb/d alpha}, \eqref{dXc/d alpha}, \eqref{dY/d alpha}, \eqref{dZa/d alpha}, \eqref{dZb/d alpha}, and \eqref{dZc/d alpha}, we obtain
\begin{align}
\frac{dX_b}{d\alpha}=&~X_b\left[-3+(X_b)^2+(X_c)^2+X_b X_c+\frac{Y^2}{2}+\frac{(Z_a)^2}{3}+\frac{(Z_b)^2}{3}+\frac{(Z_c)^2}{3}\right]+\frac{(Z_a)^2}{3}-\frac{2(Z_b)^2}{3}+\frac{(Z_c)^2}{3},\label{autonomous eq Xy}\\
\frac{dX_c}{d\alpha}=&~X_c\left[-3+(X_b)^2+(X_c)^2+X_b X_c+\frac{Y^2}{2}+\frac{(Z_a)^2}{3}+\frac{(Z_b)^2}{3}+\frac{(Z_c)^2}{3}\right]+\frac{(Z_a)^2}{3}+\frac{(Z_b)^2}{3}-\frac{2(Z_c)^2}{3},\label{autonomous eq Xz}\\
\frac{dY}{d\alpha}=&~\left(\lambda+Y \right)\left[-3+(X_b)^2+(X_c)^2+X_b X_c+\frac{Y^2}{2}+\frac{(Z_a)^2}{3}+\frac{(Z_b)^2}{3}+\frac{(Z_c)^2}{3}\right]\nonumber\\
&+\left(\frac{\lambda}{6}+\rho_a\right)(Z_a)^2+\left(\frac{\lambda}{6}+\rho_b\right)(Z_b)^2+\left(\frac{\lambda}{6}+\rho_c\right)(Z_c)^2,\label{autonomous eq Y}\\
\frac{dZ_a}{d\alpha}=&~Z_a\left[-2-X_b-X_c-\rho_a Y+(X_b)^2+(X_c)^2+X_b X_c+\frac{Y^2}{2}+\frac{(Z_a)^2}{3}+\frac{(Z_b)^2}{3}+\frac{(Z_c)^2}{3}\right],\label{autonomous eq Zx}\\
\frac{dZ_b}{d\alpha}=&~Z_b\left[-2+X_b-\rho_{b} Y+(X_b)^2+(X_c)^2+X_b X_c+\frac{Y^2}{2}+\frac{(Z_a)^2}{3}+\frac{(Z_b)^2}{3}+\frac{(Z_c)^2}{3}\right],\label{autonomous eq Zy}\\
\frac{dZ_c}{d\alpha}=&~Z_c\left[-2+X_c-\rho_{c} Y+(X_b)^2+(X_c)^2+X_b X_c+\frac{Y^2}{2}+\frac{(Z_a)^2}{3}+\frac{(Z_b)^2}{3}+\frac{(Z_c)^2}{3}\right].\label{autonomous eq Zz}
\end{align}
We now solve the set of equations \eqref{fixed point system of eqs} to determine the corresponding fixed points. 

For $Z_a=Z_b=Z_c=0$, the set of equations \eqref{fixed point system of eqs} implies
\begin{align}
X_b\left[-3+(X_b)^2+(X_c)^2+X_b X_c+\frac{Y^2}{2}\right]&=0,\label{fixed point type 0 eq 1}\\
X_c\left[-3+(X_b)^2+(X_c)^2+X_b X_c+\frac{Y^2}{2}\right]&=0,\label{fixed point type 0 eq 2}\\
(Y+\lambda)\left[-3+(X_b)^2+(X_c)^2+X_b X_c+\frac{Y^2}{2}\right]&=0.\label{fixed point type 0 eq 3}
\end{align}
With the condition \eqref{Hamiltonian constraint}, Eqs. \eqref{fixed point type 0 eq 1}, \eqref{fixed point type 0 eq 2}, and \eqref{fixed point type 0 eq 3} can be solved to give the corresponding fixed point type 0 \eqref{fixed point type 0}.

For $Z_a\neq 0$ and $Z_b=Z_c=0$, the set of equations \eqref{fixed point system of eqs} reduces to
\begin{align}
X_b\left[-3+(X_b)^2+(X_c)^2+X_b X_c+\frac{Y^2}{2}+\frac{(Z_a)^2}{3}\right]+\frac{(Z_a)^2}{3}&=0,\label{fixed point type Ia eq 1}\\
X_c\left[-3+(X_b)^2+(X_c)^2+X_b X_c+\frac{Y^2}{2}+\frac{(Z_a)^2}{3}\right]+\frac{(Z_a)^2}{3}&=0,\label{fixed point type Ia eq 2}\\
(\lambda+Y)\left[-3+(X_b)^2+(X_c)^2+X_b X_c+\frac{Y^2}{2}+\frac{(Z_a)^2}{3}\right]+\left(\frac{\lambda}{6}+\rho_a\right)(Z_a)^2 &=0,\label{fixed point type Ia eq 3}\\
-2-X_b-X_c-\rho_a Y+(X_b)^2+(X_c)^2+X_b X_c+\frac{Y^2}{2}+\frac{(Z_a)^2}{3}&=0.\label{fixed point type Ia eq 4}
\end{align}
From Eqs. \eqref{fixed point type Ia eq 1} and \eqref{fixed point type Ia eq 2}, we obtain
\begin{align}
X_c=X_b.\label{type Ia Xc in terms of Xb}
\end{align}
Consequently, Eqs. \eqref{fixed point type Ia eq 2} and \eqref{fixed point type Ia eq 3} become
\begin{align}
X_b\left[-3+3(X_b)^2+\frac{Y^2}{2}+\frac{(Z_a)^2}{3}\right]+\frac{(Z_a)^2}{3}&=0,\\
(\lambda +Y)\left[-3+3(X_b)^2+\frac{Y^2}{2}+\frac{(Z_a)^2}{3}\right]+\left(\frac{\lambda}{6}+\rho_a\right)(Z_a)^2&=0,
\end{align}
respectively, from which we can express $(Z_a)^2$ and $Y$ in terms of $X_b$ as
\begin{align}
Y&=\left(\frac{\lambda}{2}+3\rho_a\right)X_b-\lambda,\label{type Ia Y in terms of Xb}\\
\frac{(Z_a)^2}{3}&=(1+\lambda\rho_a)X_b -\left(2+\frac{\lambda\rho_a}{2}+3\rho_a^2\right)X_b^2.\label{type Ia Za in terms of Xb}
\end{align}
Inserting Eqs. \eqref{type Ia Xc in terms of Xb}, \eqref{type Ia Y in terms of Xb}, and \eqref{type Ia Za in terms of Xb} into Eq. \eqref{fixed point type Ia eq 4}, we get a quadratic equation of $X_b$, which admits two solutions,
\begin{align}
X_b&=\frac{2(-4+\lambda^2+2\lambda\rho_a)}{8+\lambda^2+8\lambda\rho_a+12\rho_a^2},\label{type Ia Xb 1}\\
X_b&=2.\label{type Ia Xb 2}
\end{align}
Putting Eq. \eqref{type Ia Xb 1} back into Eqs. \eqref{type Ia Xc in terms of Xb}, \eqref{type Ia Y in terms of Xb}, and \eqref{type Ia Za in terms of Xb}, we obtain the fixed point type $\text{I}_a$ \eqref{fixed point type Ia}. We ignore the solution \eqref{type Ia Xb 2} since it leads to $(Z_a)^2<0$. It should be noted that the derivations of the fixed points type $\text{I}_b$ and $\text{I}_c$ can be achieved similarly.

For $Z_a=0$, $Z_b\neq 0$ and $Z_c \neq 0$, the system \eqref{fixed point system of eqs} becomes
\begin{align}
X_b\left[-3+(X_b)^2+(X_c)^2+X_b X_c+\frac{Y^2}{2}+\frac{(Z_b)^2}{3}+\frac{(Z_c)^2}{3}\right]-\frac{2(Z_b)^2}{3}+\frac{(Z_c)^2}{3}&=0,\label{fixed point type IIbc eq 1}\\
X_c\left[-3+(X_b)^2+(X_c)^2+X_b X_c+\frac{Y^2}{2}+\frac{(Z_b)^2}{3}+\frac{(Z_c)^2}{3}\right]+\frac{(Z_b)^2}{3}-\frac{2(Z_c)^2}{3}&=0,\label{fixed point type IIbc eq 2}\\
(\lambda +Y)\left[-3+(X_b)^2+(X_c)^2+X_b X_c+\frac{Y^2}{2}+\frac{(Z_b)^2}{3}+\frac{(Z_c)^2}{3}\right]+\left(\frac{\lambda}{6}+\rho_b\right)(Z_b)^2+\left(\frac{\lambda}{6}+\rho_c\right)(Z_c)^2&=0,\label{fixed point type IIbc eq 3}\\
-2+X_b-\rho_{b} Y+(X_b)^2+(X_c)^2+X_b X_c+\frac{Y^2}{2}+\frac{(Z_b)^2}{3}+\frac{(Z_c)^2}{3}&=0,\label{fixed point type IIbc eq 4}\\
-2+X_c-\rho_{c} Y+(X_b)^2+(X_c)^2+X_b X_c+\frac{Y^2}{2}+\frac{(Z_b)^2}{3}+\frac{(Z_c)^2}{3}&=0.\label{fixed point type IIbc eq 5}
\end{align}
From Eqs. \eqref{fixed point type IIbc eq 4} and \eqref{fixed point type IIbc eq 5}, we obtain
\begin{align}
X_c=X_b+(\rho_c-\rho_b)Y.\label{type IIbc Xc in terms of Xb and Y}
\end{align}
In addition, Eqs. \eqref{fixed point type IIbc eq 1}, \eqref{fixed point type IIbc eq 2}, \eqref{fixed point type IIbc eq 4}, and \eqref{type IIbc Xc in terms of Xb and Y} can be solved to obtain the corresponding expressions for $(Z_b)^2$ and $(Z_c)^2$ given by
\begin{align}
(Z_b)^2&=(1+X_b-\rho_b Y)\left[-3X_b+(\rho_b-\rho_c)Y\right],\label{type IIbc Zb in terms of Xb and Y}\\
(Z_c)^2&=(1+X_b-\rho_b Y)\left[-3X_b+2(\rho_b-\rho_c)Y\right],\label{type IIbc Zc in terms of Xb and Y}
\end{align}
respectively. Then, by inserting  Eqs. \eqref{fixed point type IIbc eq 4}, \eqref{type IIbc Zb in terms of Xb and Y}, and \eqref{type IIbc Zc in terms of Xb and Y} into Eq. \eqref{fixed point type IIbc eq 3}, we obtain a quadratic equation for $Y$ with two solutions given by
\begin{align}
Y&=\frac{2\left[\lambda+(\lambda+3\rho_b +3\rho_c)X_b\right]}{-2+\lambda(\rho_b -\rho_c)+2\rho_b^2 -4\rho_c^2+2\rho_b \rho_c},\label{type IIbc Y in terms of Xb 1}\\
Y&=\frac{1+X_b}{\rho_b}. \label{type IIbc Y in terms of Xb 2}
\end{align}
It turns out that the solution \eqref{type IIbc Y in terms of Xb 2} must be ignored since  it violates the constraint \eqref{Hamiltonian constraint}. Inserting Eqs. \eqref{type IIbc Xc in terms of Xb and Y}, \eqref{type IIbc Zb in terms of Xb and Y}, \eqref{type IIbc Zc in terms of Xb and Y}, and \eqref{type IIbc Y in terms of Xb 1} into Eq. \eqref{fixed point type IIbc eq 4}, we obtain a quadratic equation of $X_b$ that has two solutions defined as
\begin{align}
X_b&=\frac{4-\lambda^2-2\lambda(2\rho_b-\rho_c)-4(\rho_b^2-2\rho_c^2+\rho_b\rho_c)}{2+\lambda^2+4\left[\lambda(\rho_b+\rho_c)+\rho_b^2+\rho_c^2+\rho_b\rho_c\right]},\label{type IIbc Xb 1}\\
X_b&=\frac{2(-1+\rho_b^2-2\rho_c^2+\rho_b \rho_c)}{2+(\rho_b-\rho_c)^2}.\label{type IIbc Xb 2}
\end{align}
Putting the solution \eqref{type IIbc Xb 1} back into Eqs. \eqref{type IIbc Xc in terms of Xb and Y}, \eqref{type IIbc Zb in terms of Xb and Y}, \eqref{type IIbc Zc in terms of Xb and Y}, and \eqref{type IIbc Y in terms of Xb 1}, we obtain the fixed point type $\text{II}_{bc}$ \eqref{fixed point type IIbc}. On the other hand, the other solution \eqref{type IIbc Xb 2} will be neglected since it violates the constraint \eqref{Hamiltonian constraint}. It is noted that one can easily apply the same procedure to obtain the fixed points type $\text{II}_{ac}$ and $\text{II}_{ab}$.

For $Z_a \neq 0$, $Z_b\neq 0$, as well as   $Z_c\neq 0$, the system \eqref{fixed point system of eqs} can be translated into the corresponding set of algebraic equations given by
\begin{align}
X_b\left[-3+(X_b)^2+(X_c)^2+X_b X_c+\frac{Y^2}{2}+\frac{(Z_a)^2}{3}+\frac{(Z_b)^2}{3}+\frac{(Z_c)^2}{3}\right]+\frac{(Z_a)^2}{3}-\frac{2(Z_b)^2}{3}+\frac{(Z_c)^2}{3}&=0,\label{fixed point type III eq 1}\\
X_c\left[-3+(X_b)^2+(X_c)^2+X_b X_c+\frac{Y^2}{2}+\frac{(Z_a)^2}{3}+\frac{(Z_b)^2}{3}+\frac{(Z_c)^2}{3}\right]+\frac{(Z_a)^2}{3}+\frac{(Z_b)^2}{3}-\frac{2(Z_c)^2}{3}&=0,\label{fixed point type III eq 2}\\
(\lambda +Y)\left[-3+(X_b)^2+(X_c)^2+X_b X_c+\frac{Y^2}{2}+\frac{(Z_a)^2}{3}+\frac{(Z_b)^2}{3}+\frac{(Z_c)^2}{3}\right]&\\
+\left(\frac{\lambda}{6}+\rho_a\right)(Z_a)^2+\left(\frac{\lambda}{6}+\rho_b\right)(Z_b)^2+\left(\frac{\lambda}{6}+\rho_c\right)(Z_c)^2 &=0,\label{fixed point type III eq 3}\\
-2-X_b-X_c-\rho_a Y+(X_b)^2+(X_c)^2+X_b X_c+\frac{Y^2}{2}+\frac{(Z_a)^2}{3}+\frac{(Z_b)^2}{3}+\frac{(Z_c)^2}{3}&=0,\label{fixed point type III eq 4}\\
-2+X_b-\rho_{b} Y+(X_b)^2+(X_c)^2+X_b X_c+\frac{Y^2}{2}+\frac{(Z_a)^2}{3}+\frac{(Z_b)^2}{3}+\frac{(Z_c)^2}{3}&=0,\label{fixed point type III eq 5}\\
-2+X_c-\rho_{c} Y+(X_b)^2+(X_c)^2+X_b X_c+\frac{Y^2}{2}+\frac{(Z_a)^2}{3}+\frac{(Z_b)^2}{3}+\frac{(Z_c)^2}{3}&=0.\label{fixed point type III eq 6}
\end{align}
As a result, by using Eqs. \eqref{fixed point type III eq 4}, \eqref{fixed point type III eq 5}, and \eqref{fixed point type III eq 6}, we can express $X_b$ and $X_c$ in terms of $Y$ as follows
\begin{align}
X_b&=\frac{-Y(\rho_a-2\rho_b+\rho_c)}{3},\label{type III Xb in terms of Y}\\
X_c&=\frac{-Y(\rho_a+\rho_b-2\rho_c)}{3}.\label{type III Xc in terms of Y}
\end{align}
Furthermore, by inserting Eqs. \eqref{type III Xb in terms of Y} and \eqref{type III Xc in terms of Y} into Eqs. \eqref{fixed point type III eq 1}, \eqref{fixed point type III eq 2}, and \eqref{fixed point type III eq 4}, we obtain the corresponding value of $(Z_a)^2$, $(Z_b)^2$, and $(Z_c)^2$ defined in terms of $Y$ as
\begin{align}
(Z_a)^2&=\frac{36-6\left[\rho_a-2(\rho_b +\rho_c)\right]Y-\left[9+2\rho_a^2 +8(\rho_b^2+\rho_c^2)-8\rho_a(\rho_b+\rho_c)-2\rho_b \rho_c\right]Y^2}{18},\label{type III Za in terms of Y}\\
(Z_b)^2&=\frac{36-6\left[\rho_b-2(\rho_a +\rho_c)\right]Y-\left[9+2\rho_b^2 +8(\rho_a^2+\rho_c^2)-8\rho_b(\rho_a+\rho_c)-2\rho_a \rho_c\right]Y^2}{18},\label{type III Zb in terms of Y}\\
(Z_c)^2&=\frac{36-6\left[\rho_c-2(\rho_a +\rho_b)\right]Y-\left[9+2\rho_c^2 +8(\rho_a^2+\rho_b^2)-8\rho_c(\rho_a+\rho_b)-2\rho_a \rho_b\right]Y^2}{18}.\label{type III Zc in terms of Y}
\end{align}
Finally, plugging Eqs. \eqref{type III Xb in terms of Y}, \eqref{type III Xc in terms of Y}, \eqref{type III Za in terms of Y}, \eqref{type III Zb in terms of Y}, and \eqref{type III Zc in terms of Y} into Eq. \eqref{fixed point type III eq 3} leads to a quadratic equation for $Y$, which can be solved to obtain two solutions given by
\begin{align}
Y&=\frac{-12}{3\lambda+2(\rho_a+\rho_b+\rho_c)},\label{type III Y 1}\\
Y&=\frac{6(\rho_a+\rho_b+\rho_c)}{3+2(\rho_a^2+\rho_b^2+\rho_c^2-\rho_a\rho_b-\rho_a\rho_c-\rho_b\rho_c)}.\label{type III Y 2}
\end{align}
As a result, the fixed point type III \eqref{fixed point type III} can be figured out by putting the solution \eqref{type III Y 1} back into Eqs. \eqref{type III Xb in terms of Y}, \eqref{type III Xc in terms of Y}, \eqref{type III Za in terms of Y}, \eqref{type III Zb in terms of Y}, and \eqref{type III Zc in terms of Y}. Apparently,  the other solution \eqref{type III Y 2} is discarded since it does not satisfy the constraint \eqref{Hamiltonian constraint}.
\section{Finding stability regions using the Routh-Hurwitz criterion} \label{RH criterion}
The entries of matrix $M$ in \eqref{perturbed dynamical system} are given by
\begin{equation}
\begin{aligned}
M_{11}&=-3+3(X_b)^2+(X_c)^2+2 X_b X_c+\frac{Y^2}{2}+\frac{(Z_a)^2}{3}+\frac{(Z_b)^2}{3}+\frac{(Z_c)^2}{3},\quad M_{12}=(X_b)^2+2X_b X_c,\\
M_{13}&=X_b Y,\quad M_{14}=\frac{2Z_a}{3}(X_b+1),\quad M_{15}=\frac{2Z_b}{3}(X_b-2),\quad M_{16}=\frac{2Z_c}{3}(X_b+1),\\ 
M_{21}&=(X_c)^2+2X_b X_c,\quad M_{22}=-3+(X_b)^2+3(X_c)^2+2X_b X_c+\frac{Y^2}{2}+\frac{(Z_a)^2}{3}+\frac{(Z_b)^2}{3}+\frac{(Z_c)^2}{3},\\
M_{23}&=X_c Y,\quad M_{24}=\frac{2Z_a}{3}(X_c+1),\quad M_{25}=\frac{2Z_b}{3}(X_c+1),\quad M_{26}=\frac{2Z_c}{3}(X_c-2),\\ 
M_{31}&=(\lambda+Y)(2X_b+X_c),\quad M_{32}=(\lambda+Y)(X_b+2X_c),\\
M_{33}&=-3+(X_b)^2+(X_c)^2+X_b X_c+\frac{3Y^2}{2}+\lambda Y+\frac{(Z_a)^2}{3}+\frac{(Z_b)^2}{3}+\frac{(Z_c)^2}{3},\\ 
M_{34}&=\left(\lambda+2\rho_a+\frac{2Y}{3}\right)Z_a,\quad M_{35}=\left(\lambda+2\rho_b+\frac{2Y}{3}\right)Z_b,\quad M_{36}=\left(\lambda+2\rho_c+\frac{2Y}{3}\right)Z_c,\\ 
M_{41}&=(-1+2X_b+X_c)Z_a,\quad M_{42}=(-1+X_b+2X_c)Z_a,\quad M_{43}=(Y-\rho_a)Z_a,\\
M_{44}&=-2-X_b-X_c-\rho_a Y+(X_b)^2+(X_c)^2+X_b X_c+\frac{Y^2}{2}+(Z_a)^2+\frac{(Z_b)^2}{3}+\frac{(Z_c)^2}{3},\\ 
M_{45}&=\frac{2Z_a Z_b}{3},\quad M_{46}=\frac{2Z_a Z_c}{3},\quad M_{51}=(1+2X_b+X_c)Z_b,\quad M_{52}=(X_b+2X_c)Z_b,\\ 
M_{53}&=(Y-\rho_b)Z_b,\quad M_{54}=\frac{2Z_a Z_b}{3},\\ 
M_{55}&=-2+X_b-\rho_b Y+(X_b)^2+(X_c)^2+X_b X_c+\frac{Y^2}{2}+\frac{(Z_a)^2}{3}+(Z_b)^2+\frac{(Z_c)^2}{3},\\ 
M_{56}&=\frac{2 Z_b Z_c}{3},\quad M_{61}=(2X_b+X_c)Z_c,\quad M_{62}=(1+X_b+2X_c)Z_c,\\ 
M_{63}&=(Y-\rho_c)Z_c,\quad M_{64}=\frac{2Z_a Z_c}{3},\quad M_{65}=\frac{2Z_b Z_c}{3},\\ 
M_{66}&=-2+X_c-\rho_c Y+(X_b)^2+(X_c)^2+X_b X_c+\frac{Y^2}{2}+\frac{(Z_a)^2}{3}+\frac{(Z_b)^2}{3}+(Z_c)^2.
\end{aligned}
\end{equation}
We then substitute each fixed point into the eigenvalue equation \eqref{eigenvalue eq}. It is not always possible to analytically solve \eqref{eigenvalue eq}, which is a polynomial equation of degree 6. We therefore employ the Routh-Hurwitz criterion, which tells us whether all the roots have negative real parts without explicitly solving the equation. For simplicity, we only consider the case where $\lambda$, $\rho_a$, $\rho_b$, and $\rho_c$ are positive. 

First, it is useful to briefly mention the Routh-Hurwitz criterion, which can be found in Ref. \cite{Merkin}. In particular, for a polynomial equation of degree $n$, 
\begin{align} \label{B1}
a_n s^n+a_{n-1}s^{n-1}+a_{n-2}s^{n-2}+...+a_1 s+a_0=0,
\end{align}
where all coefficients are real and $a_n$ is chosen to be positive, we define the $n\times n$ Hurwitz matrix to be
\begin{align}
\mathcal{H}\equiv
\begin{pmatrix}
a_{n-1} & a_{n-3} & a_{n-5} & a_{n-7} & \cdots & 0\\
a_n & a_{n-2} & a_{n-4} & a_{n-6} & \cdots & 0\\
0 & a_{n-1} & a_{n-3} & a_{n-5} & \cdots & 0\\
0 & a_n & a_{n-2} & a_{n-4} & \cdots & 0\\
\vdots & \vdots & \vdots & \vdots & \ddots & \vdots\\
0 & 0 & 0 & 0 & \cdots & a_0
\end{pmatrix}
.
\end{align} 
Then all the roots of the equation \eqref{B1} have negative real parts if and only if all the leading principal minors of the matrix $\mathcal{H}$, defined as
\begin{align}
\Delta_1\equiv
\begin{vmatrix}
a_{n-1}
\end{vmatrix},
\quad
\Delta_2\equiv
\begin{vmatrix}
a_{n-1} & a_{n-3}\\
a_n & a_{n-2}
\end{vmatrix},
\quad
\Delta_3\equiv
\begin{vmatrix}
a_{n-1} & a_{n-3} & a_{n-5}\\
a_n & a_{n-2} & a_{n-4}\\
0 & a_{n-1} & a_{n-3}
\end{vmatrix},
\quad
\ldots,
\quad
\Delta_n\equiv
\begin{vmatrix}
\mathcal{H}
\end{vmatrix},
\end{align}
are positive. Remarkably, an equivalent way to state the Routh-Hurwitz criterion using Routh table can be found in Ref. \cite{Nise}.

We can now use the Routh-Hurwitz criterion to determine the stability regions of the fixed points. Let us start with the fixed point type 0. The corresponding eigenvalue equation is
\begin{align}
\left(s-\frac{\lambda^2}{2}+3\right)^3 \left(s-\frac{\lambda^2}{2}-\lambda\rho_a+2\right)\left(s-\frac{\lambda^2}{2}-\lambda\rho_b+2\right)\left(s-\frac{\lambda^2}{2}-\lambda\rho_c+2\right)=0.
\end{align}

It is easy to see that all the roots of the equation have negative real parts if and only if the conditions shown in Eq. \eqref{fixed point type 0 st cond} are all satisfied. Therefore, the stability region of the fixed point type 0 is truly described by the inequalities  \eqref{fixed point type 0 st cond}.

For the fixed point type $\text{I}_a$, the corresponding eigenvalue equation is
\begin{align}
&\left(\frac{s}{3}+\frac{8-\lambda^2+4\lambda\rho_a+12\rho_a^2}{8+\lambda^2+8\lambda\rho_a+12\rho_a^2}\right)^2\left[\frac{s}{6}-\frac{-4+\lambda^2+2\lambda\rho_b-4\rho_a(\rho_a-\rho_b)}{8+\lambda^2+8\lambda\rho_a+12\rho_a^2}\right]\left[\frac{s}{6}-\frac{-4+\lambda^2+2\lambda\rho_c-4\rho_a(\rho_a-\rho_c)}{8+\lambda^2+8\lambda\rho_a+12\rho_a^2}\right]\nonumber\\
&\times\left[\frac{s^2}{3}+\frac{8-\lambda^2+4\lambda\rho_a+12\rho_a^2}{8+\lambda^2+8\lambda\rho_a+12\rho_a^2}s+\frac{6(-4+\lambda^2+2\lambda\rho_a)(8-\lambda^2+4\lambda\rho_a+12\rho_a^2)(2+\lambda\rho_a+2\rho_a^2)}{(8+\lambda^2+8\lambda\rho_a+12\rho_a^2)^2}\right]=0.
\end{align}
All the roots of the equation have negative real parts if and only if
\begin{equation}
\begin{aligned}
-4+\lambda^2+2\lambda\rho_b-4\rho_a(\rho_a-\rho_b)&<0,\\
-4+\lambda^2+2\lambda\rho_c-4\rho_a(\rho_a-\rho_c)&<0,\\
-4+\lambda^2+2\lambda\rho_a&>0,\\
8-\lambda^2+4\lambda\rho_a+12\rho_a^2&>0,
\end{aligned}
\end{equation}
according to the Routh-Hurwitz criterion. However, we can remove the fourth constraint since
\begin{align}
8-\lambda^2+4\lambda\rho_a+12\rho_a^2=\left[4+2(2\rho_a+\rho_b)(\lambda+2\rho_a)\right]+\left[4-\lambda^2-2\lambda\rho_b+4\rho_a(\rho_a-\rho_b)\right]>0,
\end{align}
given the first constraint. The stability region of the fixed point type $\text{I}_a$ is therefore given by the inequalities \eqref{fixed point type Ia st cond}. The stability regions of the fixed points type $\text{I}_b$ and $\text{I}_c$ are derived similarly.

We continue with the fixed point type $\text{II}_{bc}$. The corresponding eigenvalue equation is
\begin{align}
&\left\lbrace \frac{s}{3}-\frac{-4+\lambda^2+2\lambda\rho_a-4[\rho_b^2+\rho_c^2-\rho_a(\rho_b+\rho_c)]}{2+\lambda^2+4\lambda(\rho_b+\rho_c)+4(\rho_b^2+\rho_c^2+\rho_b \rho_c)}\right\rbrace\left[\frac{s}{6}+\frac{1+\lambda(\rho_b+\rho_c)+2(\rho_b^2+\rho_c^2+\rho_b\rho_c)}{2+\lambda^2+4\lambda(\rho_b+\rho_c)+4(\rho_b^2+\rho_c^2+\rho_b \rho_c)}\right]\nonumber\\
&\times [\text{a polynomial of degree 4}]=0.
\end{align}

All the roots of the equation have negative real parts if and only if
\begin{align}\label{Vec2yz RHc 1}
-4+\lambda^2+2\lambda\rho_a-4[\rho_b^2+\rho_c^2-\rho_a(\rho_b+\rho_c)]<0,
\end{align}
and
\begin{align}\label{Vec2yz RHc 2}
&\text{all roots of the polynomial of degree 4 have negative real parts}.
\end{align}

\begin{figure}[H]
\centering
\includegraphics[width=0.35\textwidth]{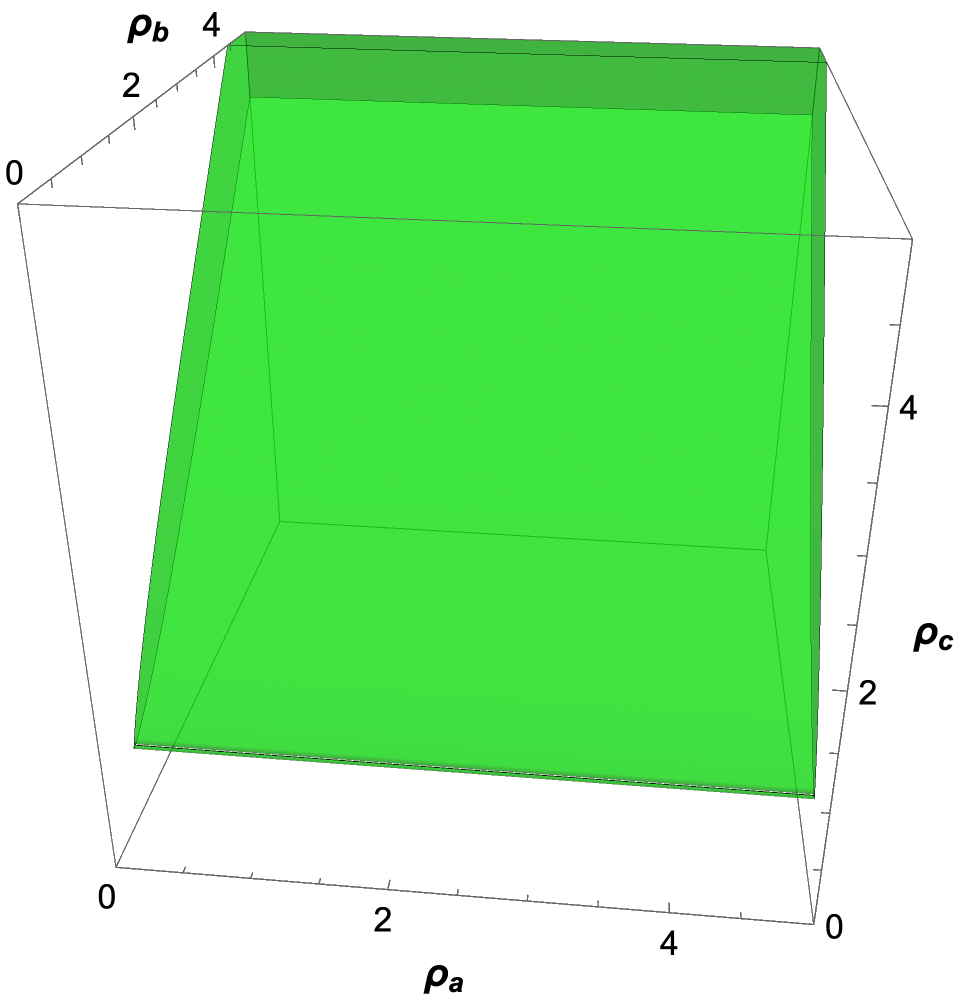}
\caption{Condition region (colored as green) for the requirement \eqref{Vec2yz RHc 2} for $\lambda=1.5$ according to the Routh-Hurwitz criterion. Note that it is very similar to Fig. \ref{fig:Vec2yzExRegion}.}
\label{fig:Vec2yzExRegionRHc}
\end{figure}

Applying the Routh-Hurwitz criterion to Eq. \eqref{Vec2yz RHc 2} is straightforward but the condition we found turns out to be very cumbersome. In principle, we can simplify the condition (for example, some constraints in the condition might be redundant and can therefore be removed) but it is not an easy task. We therefore try another approach. In particular, we plot the unsimplified stability region according to  the Routh-Hurwitz criterion in parameter space and try to guess its simplified version by comparing it to the existence region. The condition for the requirement \eqref{Vec2yz RHc 2} is depicted in Fig. \ref{fig:Vec2yzExRegionRHc}. We see that it is very similar to Fig. \ref{fig:Vec2yzExRegion}, which is the visualization of the fixed point type $\text{II}_{bc}$'s existence region \eqref{fixed point type IIbc ex cond}. After examining other values of $\lambda$, we can conclude that the condition for the requirement \eqref{Vec2yz RHc 2} is exactly described by the inequalities shown in Eq. \eqref{fixed point type IIbc ex cond}, while the stability region of the fixed point type $\text{II}_{bc}$ is displayed by Eq. \eqref{fixed point type IIbc st cond}. 

\begin{figure}[H]
\centering
\includegraphics[width=0.35\textwidth]{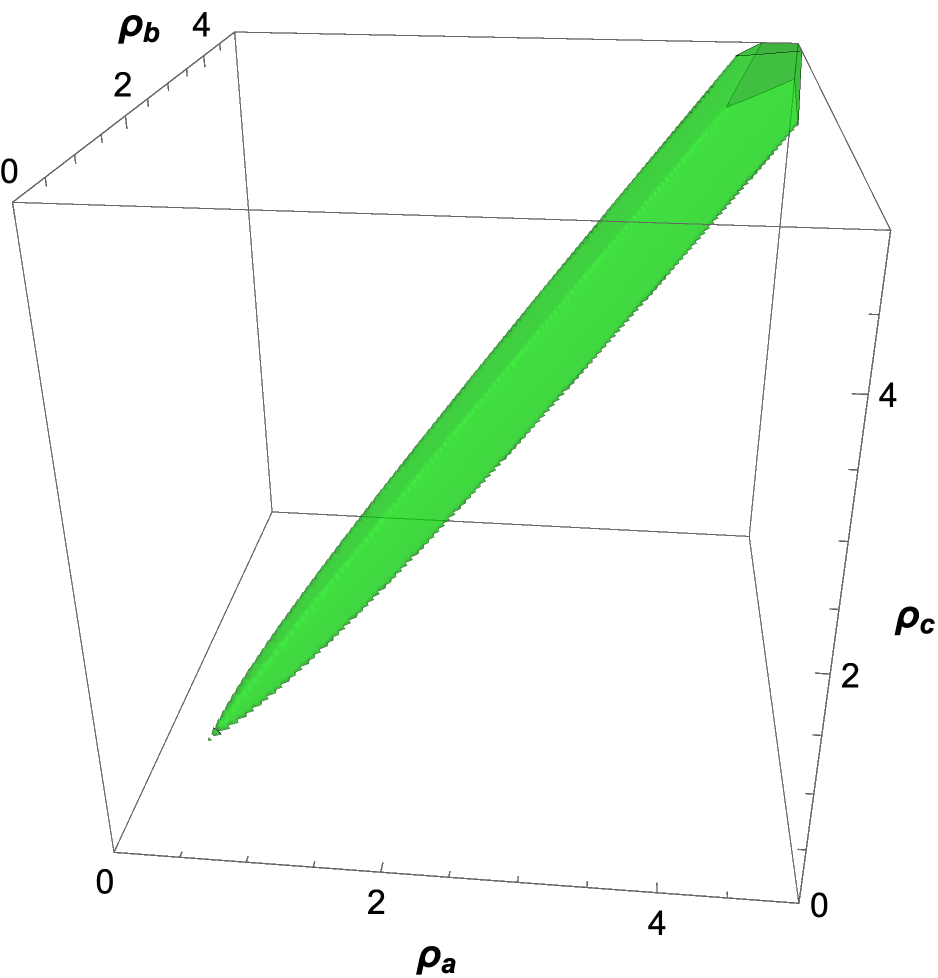}
\caption{Stability region (colored as green) of the fixed point type III for $\lambda=1.5$ according to the Routh-Hurwitz criterion.}
\label{fig:Vec3StRegionRHc}
\end{figure}

For the fixed point type III, the eigenvalue equation is a very cumbersome polynomial equation of degree 6 that is not worth writing down here. However, we can plot its stability region in Fig. \ref{fig:Vec3StRegionRHc}. We can clearly see that it is very similar to Fig. \ref{fig:Vec3StRegion}. After examining other values of $\lambda$, we can conclude that the stability region of the fixed point type III is determined by Eq. \eqref{fixed point type III ex cond}. 

%%%%%%%%%%%%%%%%%%%%%%

\end{document}